\documentclass[useAMS,usenatbib,prd,twocolumn,nofootinbib]{revtex4-1}

\def\apj{{ApJ}}                 
\def\apjl{{ApJ}}                
\def\aap{{A\&A}}                

\def\mnras{{MNRAS}}             
\def\prd{{Phys.~Rev.~D}}        
\def\prl{{Phys.~Rev.~Lett.}}    
\usepackage{amsmath,amstext}
\usepackage[T1]{fontenc}
\usepackage{amssymb}
\input{epsf}
\usepackage{graphicx}
\usepackage{ae,aecompl}
\usepackage{subfigure}
\usepackage{float}

\usepackage{hyperref}
\usepackage{amsmath}
\usepackage{amssymb}
\usepackage{mathtools}
\usepackage{bm}
\usepackage{cleveref}
\usepackage{tensor}
\usepackage{braket}

\usepackage{color}

\newcommand{\vect}[1]{\mathbf{{#1}}}

\newcommand{\spc}{\quad \quad \quad}

\def\be{\begin{equation}}
\def\ee{\end{equation}}
\def\beq{\begin{eqnarray}}
\def\eeq{\end{eqnarray}}

\begin{document}
\title{Bulk viscosity in relativistic fluids: from thermodynamics to hydrodynamics}
\author{L.~Gavassino, M.~Antonelli \& B.~Haskell}
\affiliation{Nicolaus Copernicus Astronomical Center, Polish Academy of Sciences, ul. Bartycka 18, 00-716 Warsaw, Poland}

\begin{abstract}
The approach of extended irreversible thermodynamics consists of promoting the dissipative fluxes to non-equilibrium thermodynamic variables. In a relativistic context, this naturally leads to the formulation of the theory of Israel and Stewart (1979), which is, to date, one of the most successful theories for relativistic dissipation. Although the generality of the principle makes it applicable to any dissipative fluid, a connection of the Israel-Stewart theory with microphysics has been established, through kinetic theory, only for the case of ideal quantum gases. By performing a convenient change of variables, we provide, for the case of bulk viscosity, an equivalent reformulation of the equations at the basis of extended irreversible thermodynamics. This approach maps any thermodynamic process which contributes to the bulk viscosity into a set of chemical reactions, whose reaction coordinates are abstract parameters describing the displacement from local thermodynamic equilibrium of the fluid element. We apply our new formalism to the case of the relativistic fluids, showing that the Israel-Stewart model for bulk viscosity is just the second-order expansion of a minimal model belonging to a larger class of non-perturbative theories for bulk viscosity which include the nuclear-reaction-mediated bulk viscosity of neutron star matter as a particular case. Furthermore, we show with concrete examples that our formalism provides new ways of computing the bulk viscosity directly and defines a simple prescription for constructing the Israel-Stewart model for a generic bulk-viscous fluid.
\end{abstract}

\maketitle

\section{Introduction}


The recent detection of gravitational waves from a neutron star binary inspiral, event GW170817 \citep{AbbottNS1}, together with the full range of electromagnetic emission in the following hours to months \citep{AbbottNS2}, has allowed for an unprecedented insight into the physics of hot dense matter \citep{AbbottNS3}. Future observations are likely to provide more valuable information on physics in such extreme conditions, and numerical relativity simulations will play a key role in interpreting the data.

Most simulations to date, with a few notable exceptions \citep{Duez04, SK17, SK172, Radice17}, have not included the transport properties of the matter, in the assumption that their effect is negligible on the dynamical time-scales of interest. The remnant of a neutron star merger is, however, a hot, metastable, neutron star surrounded by a thick torus, and the evolution of the system, including the ejecta and associated electromagnetic emission, will be strongly influenced by viscosity \citep{RadiceEjecta, RadiceEjecta2}. In particular the interior of the merger remnant is characterised not only by high densities but also temperatures above 10 MeV \citep{Perego19}, and in these conditions the transport time-scales are comparable to the dynamical time-scales of the system, and thermal transport and bulk viscosity play a key role in the evolution \citep{AlfordRezzolla}.

Besides its importance for applications to simulations of neutron star mergers, bulk viscosity has been shown to be a distinctive feature of fluids coupled with radiation \citep{UdeyIsrael1982,gavassino2020radiation}. It also plays an important role in heavy ion collisions \citep{FlorkowskiReview2018} and it is the only relevant dissipative process in homogeneous cosmologies \citep{Maartens1995}.

In the standard Navier-Stokes formulation, the dissipative terms of the energy-momentum tensor are taken to be proportional to the spatial derivatives of the fundamental hydrodynamic fields, temperature (for the heat flux) and velocity (for viscosity). This approach, which leads to a parabolic system \citep{Geroch95}, has been shown to lead to disastrous consequences in a relativistic framework, producing unstable \citep{Hiscock_Insatibility_first_order}, and non-causal solutions. Recent studies have also shown that the entropy of these fluids is not maximum in equilibrium \citep{GavassinoLyapunov2020}, which makes the theory inconsistent with the principles of thermodynamics and constitutes the deep origin of the instability. 

More successful attempts to model dissipation in a relativistic framework have been carried out by \citet{Stewart_1977} and \citet{Israel_Stewart_1979}. Following an approach which was later systematized by \citet{Jou_Extended}, known as \textit{extended irreversible thermodynamics}, the dissipative fluxes (e.g. viscous stresses and heat flux) are treated as further variables in the equation of state, which parametrise the displacement from equilibrium of the fluid elements. In this way one is naturally lead to write telegraph-type equations which, besides a source term containing the spatial derivatives, include a relaxation term. This encodes the delay of response of the fluid element to any displacement from equilibrium produced by the hydrodynamic motion and contains the information about the time-scale needed to the relaxation processes to restore local thermodynamic equilibrium.

Despite the many successes of this approach, which has been proven to be causal and stable in the limit of small deviations from equilibrium \citep{Hishcock1983}, still some problems remain. It has been shown to be non-causal and unstable when the deviations from equilibrium become large \citep{Hishcock1988}. This is not so surprising, considering that the formalism is a perturbative expansion near equilibrium and is not expected to hold in a more general context. On the other hand, for a philosophical desire of completeness of the theory, as well as for a more practical requirement of reliability of the numerical implementation, one would like to have a theory able, at least in principle, to deal with an arbitrary large displacement from equilibrium (provided that a hydrodynamic description remains possible). Furthermore, the contact of the Israel-Stewart theory with microscopic models has been clearly established only for ideal quantum gases, for which a full kinetic description is available. The application of the model to liquids or multifluids still constitutes a challenge \citep{andersson2020relativistic}.    

Bulk viscosity is the result of the competition between the dissipative processes which try to maintain the fluid elements in local thermodynamic equilibrium and the hydrodynamic expansion of their volume, which drives them out of it. 
In this paper we show that, if one assumes the validity of the principles of extended irreversible thermodynamics (which we summarise in subsection \ref{Qes}) and performs a convenient change of variables, any source for bulk viscosity can be modelled as a set of effective chemical reactions. This enables us to transform any Israel-Stewart \citep{Israel_Stewart_1979} bulk-viscous fluid into an effective multi-constituent single fluid.

The mathematical structure of our approach ultimately builds on the multifluid hydrodynamical formulation of Carter and collaborators \citep{noto_rel, carter1991, Carter92, carterB98}, by supplementing it with an equation of state that depends on entropy density (which allows for a causal description of heat transport, as shown e.g. in \cite{carter_proc_88, Lopez09, Lopez11}), and the `slow' degrees of freedom which give rise to bulk viscosity during an expansion. This naturally provides a general relativistic formulation for the viscous hydrodynamics which is symmetric hyperbolic and causal \citep{Causality_bulk}, and has a clear link to microphysics and kinetic theory (see also \cite{AC15} for an alternative discussion of the connection between microphysical quantities and hydrodynamics in a relativistic context).

We also provide two concrete examples, respectively for the case of bulk viscosity in neutron stars induced by $\beta$ reactions and for the case of an ideal gas of neutrons, and compare our approach to that of \citet{Israel_Stewart_1979}. In the first case we derive the formulas for the coefficients of Israel-Stewart theory (which in our formalism is simply the expansion near thermodynamic equilibrium of a more general model) in terms of thermodynamic quantities appearing in the two-component model. In the second case we find that for temperatures below 1 MeV ($\approx 10^{10}$ K) our model, which builds directly on the evolution of the momentum distributions of the particles, converges to \cite{Israel_Stewart_1979}. However at higher temperatures, such as those of a neutron star merger remnant, the results differ.

Throughout the paper we adopt the spacetime signature $ ( - , +, + , + ) $ and work in natural units $c=G=k_B=1$.


\section{Thermodynamics of out-of-equilibrium fluids}\label{TOOOEF}

In this section we extend the thermodynamic formalism to substances which are not in equilibrium. 
In general, an out-of-equilibrium system may be impossible to study without taking into account all the microscopic degrees of freedom: in the absence of an equation of state involving a limited set of macroscopic variables any hydrodynamic description of the system would be incomplete and one should rely on kinetic theory. 
Such a system would be beyond the present discussion. 
Instead, coherently with the assumptions of extended irreversible thermodynamics \cite{Jou_Extended}, we analyse systems in which it is possible to identify a limited number of macroscopic degrees of freedom containing all the information we need. 
We show that, if the volume element is isotropic, there is a thermodynamic equivalence with reacting mixtures.

\subsection{Quasi-equilibrium states}\label{Qes}

Consider a macroscopic portion of a fluid, comprised of $N$ particles, enclosed in a cubic box of volume $V=L^3$, surrounded by perfectly reflecting walls. 
We impose that $N$ is fixed (we are assuming that $N$ is a conserved charge of the underlying microscopic theory, e.g. baryon number), but allow for external variations of $V$. The system has been prepared at rest in a configuration that is homogeneous and this property will be preserved during the whole evolution. 
Therefore, the density of particles $n$ and the energy density $\mathcal{U}$ are everywhere given by
\begin{equation}\label{gandalf}
n = \dfrac{N}{V}  \spc  \mathcal{U} = \dfrac{E}{V},
\end{equation}
where $E$ is the total mass-energy of the fluid. 
Since the number of particles is fixed, it is more convenient to consider, as fundamental intensive variables, the volume and the energy per particle,  
\begin{equation}
\label{inv}
v = \dfrac{V}{N}= \dfrac{1}{n} \spc  \tilde{\mathcal{U}} = \dfrac{E}{N}=\dfrac{\mathcal{U}}{n}. 
\end{equation}
If the walls are fixed (so the volume is conserved) and adiabatic (so the energy is conserved), the system is isolated. In this case the walls play the role of external fields and appear in the microscopic Hamiltonian as parameters, not as additional degrees of freedom \citep{landau5}. Assuming that there are no other constants of motion, after an equilibration process the  fluid will reach a homogeneous equilibrium macrostate whose properties will be functions only of $V$, $N$ and $E$.

Let us assume  that the dynamics of the equilibration shows two different time-scales. In particular we impose that there is a limited set of independent observables $\alpha_A$, $A=1,...,l-1$ (the reason why we denote the amount of variables as $l-1$ will be clear soon) describing homogeneous local properties of the fluid elements, whose equilibration time $\tau_M$ is large with respect to the equilibration time $\tau_m$ of all the remaining microscopic degrees of freedom. This allows us to define a manifold $\mathcal{Z}$ of the quasi-equilibrium\footnote{
    The term quasi-equilibrium should not be confused with near-equilibrium. The first term means that all the degrees of freedom apart from a limited amount of macroscopic variables is in equilibrium. The second one means that all the degrees of freedom assume an average value which is near to the equilibrium one. A quasi-equilibrium state is also near-equilibrium when the deviation from equilibrium of the variables $\alpha_A$ is small. This is not required in the present discussion.
    } macrostates 
\begin{equation}\label{charttt}
(v, \tilde{\mathcal{U}} ,\alpha_1,...,\alpha_{l-1}).
\end{equation}  
Since our aim is to encode only bulk viscosity effects, neglecting shear viscosity and heat flow, we impose that any anisotropy equilibrates in the time-scale $\tau_m$, therefore the quasi-equilibrium states of matter elements are all invariant under rotation (i.e. isotropic).

We can introduce the entropy of the quasi-equilibrium macrostates $S$. Its amount per particle is 
\begin{equation}
x_s = \dfrac{S}{N}
\end{equation}
and is a function of the variables of state given in \eqref{charttt}. Since there are no other constants of motion, the thermodynamic equilibrium macrostate of the fluid, which is reached in a time $\tau_M$, maximizes $x_s$ compatibly with the constraints $\delta N =0$, $\delta V =0$ and $\delta E =0$, giving
\begin{equation}\label{equilllll}
\dfrac{\partial x_s}{\partial \alpha_A} \bigg|_{v,\tilde{\mathcal{U}}, \alpha_B} =0, \spc \forall A,B\neq A.
\end{equation}
The usual approach of extended irreversible thermodynamics presented by \citet{Jou_Extended} consists of choosing the bulk viscosity $\Pi$ as the unique non-equilibrium state variable $\alpha_A$, producing a model with $l=2$. This choice would lead us directly to a rediscovery of the Israel-Stewart theory for bulk viscosity \citep{Israel_Stewart_1979}. In this paper, however, we propose a more universal approach, which can be applied also to cases with $l>2$. A simple example of a system with $l=3$ is radiation hydrodynamics, in those cases in which both the photon number and the average photon energy behave as two independent (non-conserved) dynamical degrees of freedom \citep{gavassino2020radiation}. Another context in which a model with $l=3$ might be unavoidable is the hydrodynamic modelling of the bulk viscosity in holographic strongly coupled gauge theories \citep{Heller2014}.

\subsection{The chemical-like chart}\label{GEOS}

To embed the thermodynamic discussion into a hydrodynamic model we can construct a convenient global chart on $\mathcal{Z}$. 

Suppose the system has been prepared in an initial state belonging to $\mathcal{Z}$ and to move the walls of the box to produce a variation $\delta V$ of the volume in a time $\tau_{fr}$ such that
\begin{equation}
\tau_m \ll \tau_{fr} \ll \tau_M.
\end{equation}
Since this transformation is slow, the microscopic degrees of freedom have time to equilibrate instantaneously during  the process (i.e. the system moves along a curve of quasi-equilibrium states in $\mathcal{Z}$). 
This implies that there is no entropy production due to the movement of the walls  \citep{landau5}. 
On the other hand, the transformation is fast with respect to $\tau_M$, implying that the equilibration processes arising from the fact that the $\alpha_A$ can be out of equilibrium do not have time to occur. 
This transformation is adiabatic and reversible, namely
\begin{equation}
\delta S =0.
\end{equation}
Given an arbitrary point in $\mathcal{Z}$ we are able to draw a curve which crosses it (parametrised with the volume per particle $v$) that describes the states which can be reached starting from the point and making an expansion, or contraction, of the volume in the time-scale $\tau_{fr}$. 
Since the process is reversible, this curve is unique and is the same if we take any point belonging to it as the starting point. We can, thus, define a vector field $W_{fr}$ on $\mathcal{Z}$ to be the generator of the flux whose orbits are these curves. 
It satisfies
\begin{equation}\label{itregrassi}
W_{fr}(v)=1  \spc   W_{fr}(x_s)=0,
\end{equation}
where we have also used the fact that the number of particles in the box is conserved during an adiabatic expansion. 
It is always possible (see appendix \ref{STGOTFE}) to find a global chart of $\mathcal{Z}$ such that the components of $W_{fr}$ assume the form of a Kronecker delta\footnote{
    Given a smooth non-vanishing vector field on a manifold, this ``straightfication'' can always be done locally \citep{Wald}. In appendix \ref{STGOTFE} we prove that in our case a global construction is also possible.
    }.
It is evident from \eqref{itregrassi} that in such a chart one of the coordinates can be taken to be $v$, giving
\begin{equation}\label{Wad}
W_{fr} = \dfrac{\partial}{\partial v} \, ,
\end{equation}
and an other coordinate can be $x_s$. 
We construct the remaining coordinates to be dimensionless and denote them $x_A$ for $A=1,...,l-1$. Given the fact that in this chart equation \eqref{Wad} holds, we have that $W_{fr}(x_A)=0$. Hence, whatever the new internal degrees of freedom are, it is always possible to choose the $l-1$ additional variables that are conserved in fast adiabatic expansions.

Writing the energy per particle as a function of these state variables, its differential on $\mathcal{Z}$ reads
\begin{equation}\label{EOS}
d \tilde{\mathcal{U}} =\Theta dx_s + \mathcal{K} dv  - \sum_{A=1}^{l-1} \mathbb{A}^A d x_A \, ,
\end{equation}
where the symbols $\Theta$ and $\mathbb{A}^A$ bear an analogy with equilibrium thermodynamics \citep{Termo}. 
In particular, $\Theta$ represents the quasi-equilibrium generalization of the notion of temperature. 
Isolating $dx_s$ we find
\begin{equation}\label{entrppia}
dx_s = \dfrac{1}{\Theta} d\tilde{\mathcal{U}} - \dfrac{\mathcal{K}}{\Theta} d v +  \sum_{A=1}^{l-1} \dfrac{\mathbb{A}^A}{\Theta} dx_A .
\end{equation}
Comparing with \eqref{equilllll}, we find that when the thermodynamic equilibrium is reached
\begin{equation}
\mathbb{A}^A =0  \spc \forall A =1,...,l-1.
\end{equation} 
Therefore, we call the quantities $\mathbb{A}^A$ \textit{generalised affinities}.

In fact, the coordinates $x_A$ are analogous to chemical fractions (or, better, reaction coordinates). Consider a reacting multicomponent system in which the equilibration processes of the momenta of the particles are much faster than the reaction rates. Then, make an expansion which is sufficiently slow that in any instant the momenta of the particles are in their equilibrium distribution, but sufficiently fast that no reactions have time to occur. In this sense chemical fractions can be regarded as the archetype of an internal degree of freedom, to be included into the equation of state, which behaves as a frozen variable under sufficiently fast volume expansions  \cite[see e.g.][]{haensel_frozen}. 
However, we have proven that these state variables can be constructed in an arbitrary system (even in the absence of real chemical reactions). 

For example, we will explicitly consider the case of a simple gas, where there are no chemical reactions: we will identify some variables $x_A$ of this kind and use them to build a chart over the quasi-equilibrium states (this is shown in section \ref{abcdefghi} starting from the kinetic description of a simple gas).

\subsection{Effective multi-constituent equation of state}

If we define the quantities
\begin{equation}
s = n x_s \spc  n_A = n x_A ,
\end{equation}
where $s$ is the entropy per unit volume, equations \eqref{inv} and \eqref{EOS}  give
\begin{equation}\label{eos}
d \mathcal{U} = \Theta ds +  \mu dn - \sum_{A=1}^{l-1} \mathbb{A}^A d n_A \, .
\end{equation}
Here, $\mu$ is the generalization of the chemical potential to quasi equilibrium states, satisfying the condition
\begin{equation}\label{euller}
\mathcal{K} = \mathcal{U} - \Theta s - \mu n +  \sum_{A=1}^{l-1} \mathbb{A}^A  n_A .
\end{equation}
The above relation, a \textit{generalised Euler relation}, defines a Legendre transformation, implying that
\begin{equation}\label{gringooooo}
d\mathcal{K} = -s d\Theta - n d\mu + \sum_{A=1}^{l-1}   n_A d\mathbb{A}^A.
\end{equation}
The variable $\mathcal{K}$ is, therefore, the \textit{generalised grand potential density}, as it would be if the quantities $n_A$ were interpreted as densities of chemical species and the affinities as the respective chemical potentials.
When the full equilibrium is reached, $\Theta$, $\mu$ and $\mathcal{K}$ reduce respectively to the usual notions of temperature, chemical potential and grand potential density.

We have thus shown that isotropic out of equilibrium systems, under the assumption of a separation of two time-scales $\tau_M$ and $\tau_m$, have an extended equation of state which is formally identical to the one of a multi-constituent single fluid: to each abstract parameter conserved in an adiabatic expansion describing the displacement from equilibrium we can associate an effective chemical species.

\subsection{The five hydrodynamic regimes}\label{timescales!!!!}

The microscopic time-scale $\tau_m$ and the macroscopic time-scale $\tau_M$ introduced in subsection \ref{Qes} also define a separation between different hydrodynamic regimes. Depending on the time-scale $\tau_H$ of the hydrodynamic process under consideration, we can identify five distinct regimes.

\begin{itemize}
\item \textit{Kinetic Regime: $\tau_H \lesssim \tau_m$}

In this limit there is no hope to get a closed hydrodynamic system of equations because the number of independent degrees of freedom diverges. In this case a full kinetic-theory description is required.

\item \textit{Frozen Regime: $\tau_m \ll \tau_H \approx \tau_{fr} \ll \tau_M$}

In this limit the relaxation processes for the variables $x_A$ do not have time to occur. On the other hand all the remaining microscopic degrees of freedom thermalize. This limit is characterised by the conservation of the fractions $x_A$ and there is no entropy production.

\item \textit{Full-Dissipation Regime: $\tau_H \approx \tau_M$}

This is the regime of maximum dissipation, which can be seen as a chemical transfusion between the fractions $x_A$. The reason why, for $\tau_H \approx \tau_M$, the dissipation is maximal can be understood by recalling that, by definition, $\tau_M$ is the characteristic time-scale over which
the  fluid under consideration (prepared with arbitrary initial
conditions) dissipates possible initial deviations from local thermodynamic equilibrium.
Hence, the dissipative processes do not have time to occur (or are not
particularly efficient) over time-scales $\tau_H \ll \tau_M$ (but still much longer than $\tau_m$). In the
opposite limit, when the expansions/contractions of the fluid are slower
than  $\tau_M$ (i.e. $\tau_H \gg\tau_M$), the reactions are fast enough to
keep all the fractions close to their equilibrium value, making the
entropy production inevitably small. On the contrary, when $\tau_H \approx \tau_M$, the fluid is in a sort of resonant state, where the out-of-equilibrium degrees of freedom respond on the same time-scale of the macroscopic motion, leading to efficient internal dissipation.

\item \textit{Parabolic (Navier-Stokes) Regime: $\tau_H \gg \tau_M$ and }
\begin{equation}
\dfrac{\partial x_A}{\partial v}\bigg|_{x_s,\mathbb{A}^B} \longrightarrow \infty
\end{equation}

In this limit the relaxation time is small with respect to the hydrodynamic time-scale, however the expansion of the volume elements produces large deviations from local equilibrium. In this limit an expansion of the volume element produces instantaneously a displacement from local equilibrium and the affinity is proportional to the expansion.

\item \textit{Equilibrium Regime: $\tau_H \gg \tau_M$}

In this limit the relaxation processes are so fast that local thermodynamic equilibrium is always achieved, reducing the model to a perfect fluid.

\end{itemize}

Throughout the paper these regimes will be studied in more detail one by one, but we have anticipated them here to give a schematic idea of the role of the time-scales $\tau_m$ and $\tau_M$.

\subsection{Chemical gauge of the effective currents}\label{gauge}

Consider an arbitrary coordinate transformation
\begin{equation}\label{qqq}
(v,x_s,x_A) \quad \longmapsto \quad (v,x_s,y_B) ,
\end{equation}
where $A,B=1,...,l-1$. This kind of transformation represents a chemical gauge-fixing of the type presented in \cite{carter_macro_2006} and generalised in \cite{Termo}. If we write the differential of the energy per particle in the new chart we obtain
\begin{equation}
d\tilde{\mathcal{U}} = \Theta' dx_s + \mathcal{K}' dv  - \sum_{B=1}^{l-1} \mathbb{B}^B d y_B ,
\end{equation}
where the coefficients $\Theta'$, $\mathcal{K}'$ and $\mathbb{B}^B$ are related with $\Theta$, $\mathcal{K}$ and $\mathbb{A}^A$ through the relations (c.f. \eqref{EOS})
\begin{equation}\label{gaugiandotutto}
\begin{split}
& \Theta = \Theta' - \sum_{B=1}^{l-1} \mathbb{B}^B \dfrac{\partial y_B}{\partial x_s} \\
& \mathcal{K}= \mathcal{K}'-\sum_{B=1}^{l-1} \mathbb{B}^B \dfrac{\partial y_B}{\partial v} \\
& \mathbb{A}^A = \sum_{B=1}^{l-1} \mathbb{B}^B \dfrac{\partial y_B}{\partial x_A}. \\
\end{split}
\end{equation}
The partial derivatives are performed taking  $y_B$  as functions of $(v,x_s,x_A)$. Let us suppose that the $y_B$ are conserved in the adiabatic expansions described is subsection \ref{GEOS}. This makes them a choice of variables which is completely equivalent to the $x_A$. Using \eqref{Wad}, we have
\begin{equation}\label{aDDib}
W_{fr}(y_B) = \dfrac{\partial y_B}{\partial v} =0,
\end{equation}
which implies
\begin{equation}
\mathcal{K} = \mathcal{K}'.
\end{equation}
We have found that, even if there is not a unique way to define the $x_A$, this does not produce any ambiguity in the definition of $\mathcal{K}$. This has a deep physical origin, as we will see in subsection \ref{Aouoeee}. Note that there is a manifestly gauge-invariant definition for $\mathcal{K}$:
\begin{equation}\label{KkK}
\mathcal{K} = W_{fr}(\tilde{\mathcal{U}}),
\end{equation}
to prove it one only has to apply \eqref{EOS} to \eqref{Wad}.

Since the transformation \eqref{qqq} is a coordinate transformation, then the matrix 
$
\partial y_B/\partial x_A
$
is invertible, implying that all the $\mathbb{A}^A=0$ if and only if all the $\mathbb{B}^B=0$. 
This descends from the fact that in both the coordinate systems the requirement of vanishing affinities represents the condition of maximum entropy (constraining $v$ and $\tilde{\mathcal{U}}$), which holds independently of the coordinate system we choose and describes the global equilibrium state.

We finally remark  that our definition of the temperature is not invariant under the coordinate transformation. Since $\Theta$ might differ from $\Theta'$ only out of equilibrium, this should not be considered a serious problem, but only a particular case of the universal ambiguity of the non-equilibrium temperature, see also \cite{JouNonEqTemp2003}. In fact unambiguous definitions of temperature are usually obtained involving an hypothetical equilibrium of the system with an ideal heat bath, which is in contrast with the idea of a quasi-equilibrium state, see appendix \ref{statmec}.

\subsection{Kinetics of the equilibration process}\label{kinetics}

Let us finally analyse the relaxation to equilibrium of the variables $x_A$. The evolution far from equilibrium, for large affinities, may be in general complicated. Hence, even if up to now our discussion is correct also far from equilibrium,  we focus here  on the case with small $\mathbb{A}^A$.

Let us follow the evolution of the system on a time-scale $\tau_M$. Assuming the walls to be blocked, the energy and the volume do not change, thus, according to the second principle of thermodynamics, we have
\begin{equation}\label{caio}
\Theta \dfrac{d x_s}{dt} = \sum_{A=1}^{l-1} \mathbb{A}^A \dfrac{dx_A}{dt} \geq 0 ,
\end{equation} 
where we have made use of \eqref{entrppia}. Pushing forward the analogies with chemical reactions we introduce the generalised reaction rates through the formula
\begin{equation}
\dfrac{dn_A}{dt} = r_A .
\end{equation}
Since all the microscopic degrees of freedom apart from the $x_A$ have a relaxation time $\tau_m \ll \tau_M$ ($\tau_M$ is the time-scale under consideration now), in each instant all the macroscopic properties of the fluid are given once the point in $\mathcal{Z}$ is known, so that we can impose $r_A=r_A(n,\Theta,\mathbb{A}^B)$. Near equilibrium we can expand to the first order $r_A$ for small affinities. Since $r_A(\mathbb{A}^B =0)=0$, we find
\begin{equation}\label{tizio}
\dfrac{dx_A}{dt} = \dfrac{1}{n} \sum_{B=1}^{l-1} \Xi_{AB} \mathbb{A}^B,
\end{equation}
where the $(l-1)\times (l-1)$ coefficients 
\begin{equation}
\Xi_{AB} = \dfrac{\partial r_A}{\partial \mathbb{A}^B} \bigg|_{\mathbb{A}^B=0}
\end{equation}
 are functions only of $n$ and $\Theta$.
 According to Onsager's principle \citep{Onsager_1931,Onsager_Casimir},
\begin{equation}
\Xi_{AB} = \Xi_{BA},
\end{equation}
therefore only
$
{l(l-1)}/{2}
$
independent coefficients must be computed with the aid of kinetic theory. 
Plugging \eqref{tizio} in \eqref{caio} we find
\begin{equation}\label{ragazzi}
\Theta \dfrac{d s}{dt} = \sum_{A,B=1}^{l-1} \Xi_{AB} \mathbb{A}^A \mathbb{A}^B  \geq 0,
\end{equation}
which must be true for any small value of the $\mathbb{A}^A$. This implies that $\Xi_{AB}$ has to be definite non-negative. However, accounting for the fact that there are no other constants of motion apart from $N$, $V$ and $E$, which is a result of the so called \textit{ergodic assumption} \citep{Parisi_book_1988}, we can replace in the above equation the $\geq$ with $>$, implying that $\Xi_{AB}$ is also invertible (therefore positive definite), producing the notable constraints
\begin{equation}
\Xi_{AA} > 0 , \spc \Xi_{AA}\Xi_{BB} > \Xi_{AB}^2 \spc  \forall A \neq B.
\end{equation}

Once the extended equation of state for quasi-equilibrium states is given through adapted equilibrium statistical mechanical calculations and the coefficients $\Xi_{AB}$ are computed in the context of kinetic theory,\footnote{
   There is no way to compute the $\Xi_{AB}$ by means of thermodynamic calculations only as thermodynamics does not study the evolution of systems, but only the properties of their macrostates seen as stationary or quasi-stationary.
} all the macroscopic properties of the system are known and it is possible to study the whole thermodynamic evolution.

In the special case in which the substance is a multi-constituent fluid and the bulk viscosity is 
due to the presence of chemical reactions, the $x_A$ are reaction coordinates and $\mathbb{A}^A$ are reaction affinities. In this case the $r_A$ are the usual reaction rates and $\Xi_{AB}$ are their first order expansion coefficients around equilibrium \citep{carter1991,Haskell2012Onsager,Termo}.  

%

\section{Dissipative hydrodynamics of locally isotropic fluids}\label{DHOLIF}

We are now ready to develop the most general hydrodynamic description of a locally isotropic out-of-equilibrium fluid consistent with the principles of extended irreversible thermodynamics.

Let us assume that we can construct a current 
\begin{equation}\label{111}
n^\nu = n u^\nu,
\end{equation} 
where $u$ is the  local four-velocity and $n$ is the rest-frame particle density, and that the continuity equation
\begin{equation}\label{continuity}
\nabla_\nu n^\nu =0
\end{equation} 
holds. The four-velocity is normalized as $u_\nu u^\nu =-1$, so that the acceleration $a^\nu = u^\rho \nabla_\rho u^\nu$ of the fluid elements satisfies  $a_\nu u^\nu =0$.

In general it is possible to construct an energy-momentum tensor for the fluid, which in the absence of other fields appears in the right-hand side of the Einstein equations,
\begin{equation}
G_{\nu \rho} = 8\pi T_{\nu \rho},
\end{equation}
where $G_{\nu \rho}$ is the Einstein tensor, so that
\begin{equation}\label{energyconservation}
\nabla_\rho T\indices{^\rho _\nu} =0.
\end{equation}
Now, the key assumption we make is that in the comoving frame locally defined by $u$ the matter element is isotropic. Hence, the energy-momentum tensor takes the form
\begin{equation}\label{perfectfluid}
T^{\nu \rho} = \mathcal{U} u^\nu u^\rho + \Psi h^{\nu \rho},
\end{equation}
where $\mathcal{U}$ and $\Psi$ are the energy density and the diagonal term of the stress tensor measured in the frame of $u$ and
\begin{equation}
h_{\nu \rho} := g_{\nu \rho} + u_\nu  u_\rho .
\end{equation}
The energy-momentum tensor has a perfect fluid form, so we call $\Psi$ generalised pressure. 
However,  at the moment no relationship between $\mathcal{U}$ and $\Psi$ is given because the system is not in thermal equilibrium. We can project \eqref{energyconservation} tangentially and orthogonally to $u$, so, using equation \eqref{perfectfluid}, we get
\begin{equation}
 \dot{\mathcal{U}} + (\mathcal{U} + \Psi)\nabla_\nu u^\nu =0 \spc a_\nu = - \dfrac{1}{\mathcal{U} + \Psi} h\indices{^\rho _\nu} \nabla_\rho \Psi , 
\end{equation}
where we have introduced the notation $\dot{f}=u^\nu \partial_\nu f$ for any function $f$. 
Using \eqref{inv} and  \eqref{continuity}, the first relation becomes
\begin{equation}\label{comparison}
\dot{\tilde{\mathcal{U}}} = -\Psi \dot{v} \, .
\end{equation}
This equation expresses the fact that since, by local isotropy, there is no heat flow, i.e. no energy flux in the frame of the fluid element
\begin{equation}
q^\nu = - T_{\rho \lambda} u^\rho h^{\lambda \nu} =0,
\end{equation}
all the variations of the energy per particle in this frame are the result of the work of $\Psi$.

\subsection{Effective multifluid hydrodynamics}\label{Aouoeee}

Let us assume that, at a kinetic level, the fluid admits the time-scale separation presented in subsection \ref{GEOS}. We define the hydrodynamic time-scale as $\tau_H = L_0/c_s$, where $L_0$ is the length-scale in which the hydrodynamic variables change and $c_s$ is the speed of sound. In this work we  always assume that $\tau_H \gg \tau_m$, otherwise a pure hydrodynamic description would not be possible. Thus the fluid is, in each point of the space-time, in local quasi-equilibrium and the fluid elements are described by the equation of state \eqref{EOS}. 
Equation  \eqref{comparison} then becomes
\begin{equation}\label{partodaqui}
(\mathcal{K}+\Psi) \dot{v} + \Theta \dot{x}_s -\sum_{A=1}^{l-1} \mathbb{A}^A \dot{x}_A =0.
\end{equation}
Now, since we are assuming that the macroscopic local properties of the fluid are completely determined once $s$, $n$ and $n_A$ are given, then an equation of state for $\Psi$ should exist. Consider an adiabatic fast expansion with time-scale $\tau_{fr}$. In this transformation both $x_s$ and $x_A$ are conserved, therefore equation \eqref{partodaqui} implies
\begin{equation}\label{grandpotepress}
\mathcal{K} = -\Psi,
\end{equation}
in full analogy with the results of equilibrium thermodynamics. With the aid of equation \eqref{gringooooo} we obtain the Gibbs-Duhem relation
\begin{equation}
d\Psi = s d\Theta + n d\mu - \sum_{A=1}^{l-1} n_A d\mathbb{A}^A  .
\end{equation}
This result means that the isotropic stresses of the fluid are completely known once an equation of state for the energy density in the frame of the fluid element is given. To obtain its formula, however, we must make a Legendre transformation also with respect to the additional state variables associated with the displacement from equilibrium of the system. We remark that equation \eqref{grandpotepress} holds only because we are using the chemical-like chart on $\mathcal{Z}$. The key ingredient, in fact, was to invoke the conservation over a time-scale $\tau_{fr}$ of $x_A$ in equation \eqref{partodaqui}, which is exactly the property the $x_A$ were designed to satisfy.

Plugging equation \eqref{grandpotepress} into \eqref{partodaqui} we find the formula
\begin{equation}\label{entttropppia}
\Theta \dot{ x}_s =\sum_{A=1}^{l-1} \mathbb{A}^A \dot{x}_A ,
\end{equation}
which describes the heat production due to the relaxation processes towards equilibrium of the variables $x_A$ and is in agreement with \eqref{caio}.

Let us now study the hydrodynamic processes which occur on a time-scale $\tau_{fr}$. At this level the transformations which the fluid elements incur are adiabatic expansions of the type described in subsection \ref{GEOS}, so, again, we can impose
\begin{equation}\label{liu}
\dot{x}_s = \dot{x}_A =0.
\end{equation}
This means that the entropy per particle and the quantities $x_A$ are frozen to a constant value along the worldline of the matter element. For this reason we can call this regime, in agreement with \citet{haensel_frozen},  \textit{frozen regime}, see subsection \ref{timescales!!!!}. In this limit the adiabatic index describing the response of the fluid to a perturbation is
\begin{equation}
\Gamma_{FR} = \dfrac{d \ln \Psi}{d \ln n} \bigg|_{x_s, x_A} .
\end{equation}
We, now, define the currents
\begin{equation}\label{222}
s^\nu = s u^\nu  \spc  n_A^\nu = n_A u^\nu,
\end{equation} 
which, using \eqref{continuity} and \eqref{liu}, are conserved:
\begin{equation}\label{333}
\nabla_\nu s^\nu =0  \spc  \nabla_\nu n_A^\nu =0.
\end{equation}
Since $\mathcal{U}$ is the energy density measure in the frame of the entropy, we call it \textit{internal energy density}. It is possible to verify that the system \eqref{111}, \eqref{continuity}, \eqref{perfectfluid}, \eqref{eos}, \eqref{grandpotepress}, \eqref{222} and \eqref{333} describes a multifluid with $l$ components locked to the entropy. Therefore, this system must arise directly from a convective variational principle and can be shown to be a well-posed problem, see \cite{andersson2007review}. 
It is also interesting to note that our result shows that in Carter's multifluid formalism the notion of \textit{current} is a very general concept, being useful to describe not only real chemical species, but also abstract non-equilibrium variables.

\subsection{Dissipative hydrodynamics}

Let us assume that $\tau_H \gtrsim \tau_{fr}$. Then, coherently with section \ref{TOOOEF}, we can also impose
\begin{equation}\label{strongG}
\dot{x}_A = \dfrac{1}{n} \sum_{B=1}^{l-1} \Xi_{AB} \mathbb{A}^B.
\end{equation}
Plugging this formula inside \eqref{entttropppia} we obtain
\begin{equation}\label{entropyprodfluid}
\Theta \dot{x}_s =\dfrac{1}{n} \sum_{A,B=1}^{l-1} \Xi_{AB} \mathbb{A}^A \mathbb{A}^B .
\end{equation}
With the aid of \eqref{continuity}, these equations can be recast in the form
\begin{equation}\label{buzzurro}
 \nabla_\nu n_A^\nu = \sum_{B=1}^{l-1} \Xi_{AB} \mathbb{A}^B \spc \Theta \nabla_\nu s^\nu =  \sum_{A,B=1}^{l-1} \Xi_{AB} \mathbb{A}^A \mathbb{A}^B . 
\end{equation}
We have shown that any dissipative process in a locally isotropic fluid can be modelled as a chemical transfusion between effective currents representing a convenient choice of coordinates in the space of quasi-equilibrium states $\mathcal{Z}$. 

Note that in the limit in which the hydrodynamic time-scale is much larger than $\tau_M$ ($\Xi_{AB} \longrightarrow + \infty$), the reactions become so fast that full thermodynamic equilibrium is everywhere achieved, namely
\begin{equation}
\mathbb{A}^A =0.
\end{equation}
This implies that in this limit
\begin{equation}
x\indices{_A} = x_A^{eq}(n,x_s)  \spc  \dot{x}_s =0.
\end{equation}
To derive the equations in this regime, one has to take the limit $\Xi_{AB} \longrightarrow + \infty$ and $\mathbb{A}^A \longrightarrow 0$ in the first equation of \eqref{buzzurro}, keeping their product finite. Regarding the second equation, since the right-hand side is quadratic in $\mathbb{A}^A$ and linear in $\Xi_{AB}$, the entropy production vanishes.

We have recovered the ideal perfect fluid and the equation of state \eqref{eos} reduces to
\begin{equation}
d\mathcal{U} = \Theta ds + \mu dn.
\end{equation}
In this limit, called \textit{equilibrium regime} (see subsection \ref{timescales!!!!}), matter reacts to perturbations with the adiabatic index (c.f. with \cite{haensel_frozen})
\begin{equation}
\begin{split}
\Gamma_{EQ}  = \, & \dfrac{d \ln \Psi (n,x_s , x_A^{eq}(n,x_s))}{d \ln n}  = \\ & \Gamma_{FR} + \dfrac{n}{\Psi} \sum_{A=1}^{l-1} \dfrac{\partial x_A^{eq}}{\partial n}\bigg|_{x_s} \dfrac{\partial \Psi}{\partial x_A} \bigg|_{n,x_s,x_B} . \\
\end{split}
\end{equation}

\subsection{Summary of the equations of the theory}

We can finally summarize the complete set of equations of the theory, valid in all the regimes (provided that $\tau_H \gg \tau_m$). The variables are $g_{\nu \rho}$, $n^\nu$, $x_s$ and all the $x_A$. The system of differential equations is
\begin{equation}
\begin{split}
& {G_{\nu \rho}}\, = \, 8 \pi (\,  n\, \tilde{\mathcal{U}} u_\nu u_\rho + \Psi h_{\nu \rho} \, )\\
& \nabla_\nu n^\nu =0 \\
&  u^\nu \nabla_\nu x_s =\dfrac{1}{n\Theta} \sum_{A,B=1}^{l-1} \Xi_{AB} \mathbb{A}^A \mathbb{A}^B \\
& u^\nu \nabla_\nu x_A = \dfrac{1}{n} \sum_{B=1}^{l-1} \Xi_{AB} \mathbb{A}^B. \\
\end{split}
\end{equation}
In addition, it is necessary to have an equation of state $\tilde{\mathcal{U}}=\tilde{\mathcal{U}}(n^{-1},x_s,x_A)$ and an expression for the transport coefficients $\Xi_{AB}=\Xi_{AB}(n,x_s)$.

In the case in which one wants to work with fixed background spacetime, then $g_{\nu \rho}$ is no more a variable and Einstein's equations can be replaced by the Euler equation
\begin{equation}
(n\tilde{\mathcal{U}}+\Psi)u^\rho \nabla_\rho u_\nu = - h\indices{^\rho _\nu} \nabla_\rho \Psi \, .
\end{equation}

\section{Emergence of bulk viscosity}\label{eobv!!!}

In this section we prove that the hydrodynamic model developed in the previous section describes, in the first order in $\mathbb{A}^A$, a locally isotropic viscous fluid, i.e. a perfect fluid with a bulk viscosity term.
We show, however, that the model has a natural hyperbolic form, being governed by a telegraph-type equation, whose relaxation time-scale is given by $\tau_M$. 
We will examine the parabolic limit for arbitrary $l$,\footnote{
    We recall that $l$ counts the number of degrees of freedom (other than $n$) that have to be included in the quasi-equilibrium equation of state together with the number density.} 
proving that the relativistic Navier-Stokes formulation is recovered. 
Finally we will summarize how the different regimes can be obtained varying the time-scales of the involved processes.

From now on we will use the Einstein summation convention for repeated indices $A,B,C,D$.

\subsection{Expansion for small deviations from equilibrium: thermodynamic potentials}\label{EFSDFETP}

It is convenient to consider the quantities per particle and use as free state variables $x_s$, $v$ and $\mathbb{A}^A$. The affinities are preferable in this context to $x_A$, because our aim is to make an expansion for small displacements from equilibrium. In this section all the quantities are assumed to be functions of these variables and the partial derivatives will be performed accordingly.

Recalling equation \eqref{EOS} and the identification \eqref{grandpotepress}, we can introduce the new thermodynamic potential
\begin{equation}\label{legendrandotogavassino}
\tilde{\mathcal{G}} := \tilde{\mathcal{U}} +  \mathbb{A}^A x_A ,
\end{equation}
whose differential is
\begin{equation}\label{dG}
d\tilde{\mathcal{G}} = \Theta dx_s - \Psi dv  +  x_A d\mathbb{A}^A .
\end{equation}
The perturbative approach is built by expanding $\tilde{\mathcal{G}}$ to second order:
\begin{equation}
\tilde{\mathcal{G}}(x_s,v,\mathbb{A}^A) = \tilde{\mathcal{U}}_{eq} + x_A^{eq}\mathbb{A}^A + \dfrac{1}{2}  \tilde{\mathcal{G}}_{AB}^{eq} \mathbb{A}^A \mathbb{A}^B,
\label{PAGURO}
\end{equation}
where the coefficients $\tilde{\mathcal{U}}_{eq}$, $x_A^{eq}$ and $\tilde{\mathcal{G}}_{AB}^{eq}$ are functions only of $x_s$ and $v$. The zeroth order term in the expansion is the energy per particle evaluated in equilibrium,
\begin{equation}
\tilde{\mathcal{U}}_{eq} =\tilde{\mathcal{U}}|_{\mathbb{A}^A =0} =\tilde{ \mathcal{G}}|_{\mathbb{A}^A =0},
\end{equation}
and the first order expansion coefficients are the equilibrium fractions
\begin{equation}
x_A^{eq} = x_A|_{\mathbb{A}^B=0} = \dfrac{\partial \tilde{\mathcal{G}}}{\partial \mathbb{A}^A} \bigg|_{\mathbb{A}^B=0}.
\end{equation}
The $(l-1)\times (l-1)$ matrix $\tilde{\mathcal{G}}_{AB}^{eq}$ in \eqref{PAGURO} is defined as
\begin{equation}
\tilde{\mathcal{G}}_{AB}^{eq} = \dfrac{\partial^2 \tilde{\mathcal{G}}}{\partial \mathbb{A}^A \partial \mathbb{A}^B} \bigg|_{\mathbb{A}^C =0}.
\end{equation}
Defining 
\begin{equation}\label{zsefvgy}
\Theta_{eq}:= \dfrac{\partial \tilde{\mathcal{U}}_{eq}}{\partial x_s}  \spc  P := -\dfrac{\partial \tilde{\mathcal{U}}_{eq}}{\partial v} 
\end{equation}
and considering equation \eqref{dG}, we obtain
\begin{equation}\label{uuuuuuu}
\begin{split}
& \Theta = \Theta_{eq} +  \mathbb{A}^A \dfrac{\partial x_A^{eq}}{\partial x_s} + \dfrac{1}{2}   \mathbb{A}^A \mathbb{A}^B \dfrac{\partial \tilde{\mathcal{G}}_{AB}^{eq}}{\partial x_s} \\
& \Psi = P - \mathbb{A}^A \dfrac{\partial x_A^{eq}}{\partial v} - \dfrac{1}{2}  \mathbb{A}^A \mathbb{A}^B \dfrac{\partial \tilde{\mathcal{G}}_{AB}^{eq}}{\partial v} \\
& x_A = x_A^{eq} + \tilde{\mathcal{G}}_{AB}^{eq} \mathbb{A}^B . \\
\end{split}
\end{equation}
The quantity $P$ is the \textit{equilibrium pressure}, the quantity 
\begin{equation}\label{qwerty}
\Pi = - \mathbb{A}^A \dfrac{\partial x_A^{eq}}{\partial v}
\end{equation}
is called \textit{first-order viscous stress} and the quantity
\begin{equation}
\Pi'' = -\dfrac{1}{2}   \mathbb{A}^A \mathbb{A}^B \dfrac{\partial \tilde{\mathcal{G}}_{AB}^{eq}}{\partial v}
\end{equation}
is called \textit{second-order viscous stress}. Therefore the second equation of \eqref{uuuuuuu} can be rewritten in the form
\begin{equation}
\Psi = P + \Pi + \Pi'',
\end{equation}
in which $P$ can be considered the thermodynamic pressure (in the sense that it can be computed directly from equilibrium thermodynamics), while $\Pi$ and $\Pi''$ constitute the first two contributions to the bulk viscosity.

Now we can reverse the Legendre transformation \eqref{legendrandotogavassino} and get
\begin{equation}\label{polkj}
\tilde{\mathcal{U}} = \tilde{\mathcal{U}}_{eq} - \dfrac{1}{2}  \tilde{\mathcal{G}}^{eq}_{AB} \mathbb{A}^A \mathbb{A}^B,
\end{equation}
so the first order correction in $\mathbb{A}^A$ to the internal energy density is zero. This formula can also be used to prove that 
\begin{equation}
\tilde{\mathcal{G}}^{eq}_{AB}     = \Theta_{eq}\dfrac{\partial^2 x_s}{\partial \mathbb{A}^A \partial \mathbb{A}^B} \bigg|_{v,\tilde{\mathcal{U}},\mathbb{A}^C=0}.
\end{equation}
Therefore the matrix $\tilde{\mathcal{G}}^{eq}_{AB}$ must be negative definite, implying that equilibrium is the minimum of the energy per particle with fixed $v$ and $x_s$, in accordance with \cite{Callen_book}. 

Up to now we have considered corrections to the second order, to keep track of the corrections induced on all the thermodynamic potentials. The first step to recover the usual formulation for the bulk viscosity is to consider the formula \eqref{perfectfluid} for the energy-momentum tensor of the system and truncate the expansion to the first order in $\mathbb{A}^A$. This leads us to
\begin{equation}
T^{\nu \rho} = \mathcal{U}_{eq} u^\nu u^\rho + (P+\Pi)h^{\nu \rho},
\end{equation}
where $\mathcal{U}_{eq}= n \tilde{\mathcal{U}}_{eq}$. We obtained the perfect fluid in local thermodynamic equilibrium, described by the equation of state $\mathcal{U}_{eq}(s,n)$, plus a bulk-viscosity correction $\Pi$ to the isotropic stresses.

\subsection{Expansion for small deviations from equilibrium: dissipation}\label{expBuba!}

We have verified that for small perturbations from equilibrium, the energy-momentum tensor takes the usual form for a bulk-viscous fluid. To complete the perturbative expansion of the theory we need to study \eqref{strongG} in the limit of small affinities. 
Let us consider $x_A$ as a function of $x_s$, $v$ and $\mathbb{A}^A$, in accordance with what we did in subsection \ref{EFSDFETP}. Then, using the chain rule, we have that equation \eqref{strongG} becomes
\begin{equation}\label{ilgonzo}
\dfrac{\partial x_A}{\partial x_s} \dot{x}_s + \dfrac{\partial x_A}{\partial v} \dot{v} +  \dfrac{\partial x_A}{\partial \mathbb{A}^B} \dot{\mathbb{A}}^B = \dfrac{1}{n}  \Xi_{AB} \mathbb{A}^B.
\end{equation}
Now, equation \eqref{continuity} implies that
\begin{equation}
\dot{v} = v \nabla_\nu u^\nu ,
\end{equation}
so, contracting \eqref{ilgonzo} with the symmetric $(l-1)\times(l-1)$ matrix $\Xi^{AB}$, defined as the inverse of $\Xi_{AB}$,
\begin{equation}\label{inversemetric}
 \Xi^{AC} \Xi_{CB} = \delta\indices{^A _B}  ,
\end{equation}
we find
\begin{equation}\label{buzzzzurro}
-n \Xi^{AB} \dfrac{\partial x_B}{\partial \mathbb{A}^C} \dot{\mathbb{A}}^C + \mathbb{A}^A - n \Xi^{AB} \dfrac{\partial x_B}{\partial x_s} \dot{x}_s = \Xi^{AB} \dfrac{\partial x_B}{\partial v} \nabla_\nu u^\nu .
\end{equation}
Now we define the $(l-1)\times (l-1)$ matrix
\begin{equation}\label{TTau}
\tau\indices{^A _C} = -n \Xi^{AB} \dfrac{\partial x_B}{\partial \mathbb{A}^C},
\end{equation}
called \textit{relaxation-time matrix}, the $l-1$ vector
\begin{equation}\label{kappa}
k^A = \Xi^{AB} \dfrac{\partial x_B}{\partial v} 
\end{equation}
and the second order $l-1$ vector, see \eqref{entropyprodfluid},
\begin{equation}
\mathcal{Q}^A = - n \Xi^{AB} \dfrac{\partial x_B}{\partial x_s} \dot{x}_s = -\Xi^{AB} \Xi_{CD} \dfrac{\partial x_B}{\partial x_s}  \dfrac{\mathbb{A}^C \mathbb{A}^D}{\Theta}  .
\end{equation} 
Therefore equation \eqref{buzzzzurro} can be rewritten in the form
\begin{equation}
\tau\indices{^A _B} \dot{\mathbb{A}}^B + \mathbb{A}^A + \mathcal{Q}^A = k^A \nabla_\nu u^\nu .
\end{equation}
The final step consists of keeping only the lowest order in the affinities, which means that the transport coefficients $\tau\indices{^A _B}$ and $k^A$ can be evaluated in $\mathbb{A}^A=0$ (thus we can put a superscript $eq$ to the quantities appearing in the right-hand side of their definitions) and $\mathcal{Q}^A$ is approximated to zero, leaving the telegraph-type equation
\begin{equation}\label{zazza}
\tau\indices{^A _B} \dot{\mathbb{A}}^B + \mathbb{A}^A = k^A \nabla_\nu u^\nu .
\end{equation}
The first term of left-hand side encodes the relaxation time-scale, clearly the matrix $\tau\indices{^A _B}$ is of first order in $\tau_M$. The right-hand side describes the fact that if the volume element expands, then the fluid is driven out of local thermodynamic equilibrium if the relaxation processes are not sufficiently fast. In particular, $k^A$ quantifies the response of the affinity $\mathbb{A}^A$ to the expansion.

In relativity, the need to have telegraph-type equations describing the evolution of the internal degrees of freedom of the matter elements is well established \citep{Hiscock_Insatibility_first_order, andersson2007review, rezzolla_book, Lopez11}. However, the relaxation term, usually inserted ad-hoc to make the theories hyperbolic\footnote{See, e.g., \cite{Kost2000} for an
introduction to hyperbolicity and parabolicity, with reference to the telegraph equation. Intuitively, the difference is in the fact that, in a parabolic system, the domain of influence of the initial data imposed on a point is bounded (locally) by a 3D hyperplane (giving rise to acausal dynamics) while, in a hyperbolic system, it is bouded by a 3D cone and can therefore be causal if this cone is contained inside the light-cone. The telegraph equation is the prototype of a dissipative hyperbolic equation. }, has been regarded by some authors more as an artefact imposed to fulfill a mathematical necessity, rather than a physically justified contribution \cite{Geroch95}.  Thus, equation \eqref{zazza}  represents a justification of the existence of this term, arising directly from arguments of non-equilibrium thermodynamics. 
Our approach, however, differs from previous ones (see \cite{Jou_Extended} for a summary), because the telegraph-type equation is derived for the more fundamental quantities $\mathbb{A}^A$ and not directly for $\Pi$.

\subsection{The parabolic limit}\label{paraparaparapara}

An important test is to verify if we can recover the usual Navier-Stokes prediction for the bulk viscosity in an appropriate limit. 

In equation \eqref{zazza} the term $\tau\indices{^A _B} \dot{\mathbb{A}}^B$ describes a delay in the response of matter to an expansion. To recover Navier-Stokes we have to assume that this term is negligible. The conditions we need to achieve our goal are the following:
\begin{equation}\label{efiijvneinvveiol}
\tau_M \longrightarrow 0  \spc  \dfrac{\partial x_A}{\partial v} \longrightarrow  \infty.
\end{equation}
In fact we need to impose that $\tau_M$ is smaller than the hydrodynamic time-scale, sufficiently short to assume that there is no long-term memory of the past, but not so small to recover the frozen regime. In particular, we need that the quantities $k^A$, given in \eqref{kappa}, remain finite. Since  
\begin{equation}
\Xi^{AB} \longrightarrow 0,
\end{equation}
we need to compensate imposing the second condition of \eqref{efiijvneinvveiol}. 

In this limit we can make the approximation
\begin{equation}\label{parabolic}
\mathbb{A}^A = k^A \nabla_\nu u^\nu .
\end{equation}
Then, using \eqref{qwerty}, we obtain
\begin{equation}
\Pi = -\zeta \nabla_\nu u^\nu ,
\end{equation}
with
\begin{equation}\label{zetta}
\zeta =\Xi^{AB} \dfrac{\partial x_A^{eq}}{\partial v} \dfrac{\partial x_B^{eq}}{\partial v}.
\end{equation}
Since $\Xi_{AB}$ is positive definite, so is $\Xi^{AB}$, and therefore
\begin{equation}
\zeta \geq 0.
\end{equation}
We conclude analysing the entropy production in the parabolic limit. If we plug \eqref{parabolic} in \eqref{entropyprodfluid} we find
\begin{equation}
\Theta \dot{x}_s = \dfrac{1}{n} \Xi_{AB} k^A k^B (\nabla_\nu u^\nu )^2.
\end{equation} 
This can be recast into the best known formula
\begin{equation}\label{sbluearf}
\Theta \nabla_\nu s^\nu = -\Pi \nabla_\nu u^\nu = \zeta \bigg( \dfrac{\dot{v}}{v} \bigg)^2 \geq 0.
\end{equation}
Therefore we have recovered all the equations of the relativistic Navier-Stokes model, which is a parabolic system, see \cite{Geroch95}. We are using the general term relativistic Navier-Stokes model to denote the \textit{first order theories} of \citet{Eckart40}, \citet{landau6} and more in general those considered by \citet{Hiscock_Insatibility_first_order}. They all converge to the unique model we presented above in the case of bulk-viscous substances, in the absence of heat flux and shear viscosity.

We remark that in the hyperbolic regime, as $l$ grows, the model becomes increasingly complicated because all the out-of-equilibrium variables have their own equilibration time and can influence the evolution of each other. 
On the other hand, in the parabolic limit  all the contributions add up in a unique factor $\zeta$ in which it is not possible to distinguish the individual microphysical processes. 


We conclude this section by expanding on the physical
meaning of the limits in \eqref{efiijvneinvveiol}. The limit $\tau_M \longrightarrow 0$ is a requirement of slow evolution of the system. This condition is also commonly used to derive the Navier-Stokes equations directly from kinetic theory \citep{huang_book}. Since the instability of relativistic Navier-Stokes arises from the existence of fast-growing unphysical modes \citep{Hiscock_Insatibility_first_order,Kost2000,GavassinoLyapunov2020}, the condition of slow evolution guarantees that the telegraph-type equation \eqref{zazza} is equivalent to \eqref{parabolic} only for physical solutions of the latter. In contrast, the theory behaves differently along the gapped modes of the system, where first-order theories become unphysical \citep{Kovtun2019} and the relaxation term $\tau\indices{^A _B} \dot{\mathbb{A}}^B$ cannot be neglected.

The second limit of \eqref{efiijvneinvveiol} is a formal way of stating that, for bulk viscosity to survive the slow limit, some $x_A^{eq}(v,x_s)$ need to have a strong dependence on $v$: they should vary considerably as the volume elements expand, forcing the fluid to dissipate energy in the attempt to re-equilibrate the fractions $x_A$ to the value $x_A^{\text{eq}}$. Only in this case the entropy production \eqref{sbluearf} is not negligible, even if it is a second-order infinitesimal (due to the slow limit applied to the expansion rate $\dot{v}/v$).

\section{Heat production in small oscillations}

In this section we study the damping of oscillations of homogeneous systems produced by a bulk viscosity term.
We show that, as our system is hyperbolic and described by telegraph-type equations \eqref{zazza}, it naturally produces a dependence on the frequency of the oscillations related to the delayed response of matter on a time-scale $\tau_M$, unlike the standard parabolic Navier-Stokes approach. Our theory thus generally includes this effect, which was studied by \citet{Sawyer_Bulk1989} for the specific case of bulk viscosity due to reactions in neutron stars.

\subsection{Setting the stage} 

To study a weak damping effect on small oscillations around a uniform equilibrium configuration we solve the linearised dynamics for perturbations in the non-dissipative limit (i.e. by replacing the bulk-viscous fluid with the perfect fluid that we would recover if the system was in the equilibrium regime: $\mathbb{A}^A =0$). Then, we compute separately the evolution of the affinities $\mathbb{A}^A$  (at the first-order) and the associated heat production (which is treated as a second-order correction), assuming that they have a negligible effect (small back-reaction limit) on the dynamics of the oscillations for the regimes and associated time-scales that are being considered \citep{Sawyer_Bulk1989,Haensel_Bulk_hyperons}. 

Hence, the first step consists of studying the oscillations of a perfect fluid in local thermodynamic equilibrium around a homogeneous static solution  (the independent variables are 5, so one equation is redundant),
\begin{equation}
\begin{split}
& \nabla_\nu n^\nu =0 \\
& \nabla_\nu s^\nu =0 \\
& u^\rho \nabla_\rho u_\nu = - \dfrac{1}{\mathcal{U}_{eq} + P} h\indices{^\rho _\nu} \nabla_\rho P . \\
\end{split}
\end{equation}
In addition we have the equation of state $\mathcal{U}_{eq}(n,s)$ and we work in the flat spacetime limit. We use the global inertial chart $(t,z^1,z^2,z^3)$, we assume invariance under translations in the directions $2$ and $3$ and we impose $u^2=u^3=0$. Any physical quantity $X$ is assumed to have a spacetime dependence of the form
\begin{equation}
X = X_0 + \delta X e^{i(kz^1-\omega t)} \, ,
\end{equation}
where $X_0$ is the unperturbed value and $\delta X$ is a complex amplitude of the perturbation (both are uniform and constant) all the fluctuations are encoded in the exponential. 

The system describes the propagation of sound waves in the medium:
\begin{equation}\label{zanzibar}
\delta v = -v \dfrac{k}{\omega} \delta u^1  \spc  \delta x_s =0  \spc \dfrac{\omega}{k} = c_s \, ,
\end{equation}
where 
\begin{equation}
c_s =\sqrt{ \dfrac{-v}{\mathcal{U}_{eq} + P} \dfrac{\partial P}{\partial v} \bigg|_{x_s}  } =\sqrt{\dfrac{\partial P}{\partial \mathcal{U}_{eq}} \bigg|_{x_s}}
\end{equation}
is the speed of sound. The perturbations to the affinities will be computed explicitly in the next subsections, assuming that \eqref{zanzibar} is approximately valid also in the bulk-viscous case and invoking the telegraph-type equation \eqref{zazza}. For the time being, we keep the discussion general, leaving the values of $\delta \mathbb{A}^A$ (and consequently of $\delta x_A$) undetermined. 

Now we compute the energy dissipated in a large volume $V$ and in a time $\tau_{d}$ which is assumed to be large with respect to $\omega^{-1}$ (in order to contain a large number of oscillations), but sufficiently small to neglect the back-reaction of dissipation on the oscillations.
The heat produced is 
\begin{equation}
\Delta Q = \Theta_0 \Delta S =\int_0^{\tau_d}  \int_V   \Xi_{AB} \mathbb{A}^A \mathbb{A}^B  \, d_3 z \,dt .
\label{PIZZAIOLOdiCARNEVALE}
\end{equation}
Since
\begin{equation}
\mathbb{A}^A_0 =0
\quad \text{and} \quad 
\mathbb{A}^A = \text{Re}\left[  \, \delta \mathbb{A}^A e^{i(kz^1-\omega t)} \, \right],
\end{equation}
the heat $\Delta Q $ is a second order effect in the fluctuations and we can replace $\Xi_{AB}$ with $\Xi_{AB}^0$
in \eqref{PIZZAIOLOdiCARNEVALE}.
In the integration  the oscillating terms (in space and time) give a negligible contribution, while the uniform terms factorise out. Thus, the average heat production per unit volume and time is
\begin{equation}\label{dqdt}
\dfrac{dq_{av}}{dt} := \dfrac{\Delta Q}{V \tau_d} = \dfrac{1}{2} \Xi_{AB}^0 \,     \delta\mathbb{A}^A (\delta\mathbb{A}^B)^* .
\end{equation}
Note that we should extract the real part in the right-hand side, but it is already real considering that $\Xi_{AB}^0$ is a real symmetric matrix.

\subsection{Heat production in the parabolic limit}\label{havana}

In the parabolic limit we need to perturb the equation \eqref{parabolic}, taking $n$, $s$ and $u$ to be given by \eqref{zanzibar}, i.e. solutions of the system in the non-dissipative limit. Therefore, at first order
\begin{equation}
\delta \mathbb{A}^A =-i \omega  \dfrac{\delta v}{v_0} k_0^A . 
\end{equation} 
Plugging this formula into \eqref{dqdt} we find
\begin{equation}
\dfrac{dq_{av}}{dt} = \dfrac{\omega^2 \delta v^2}{2v_0^2} \Xi_{AB}^0 k_0^A k_0^B .
\end{equation}
With the aid of \eqref{kappa} and \eqref{zetta} we rewrite the above expression in the form
\begin{equation}
\dfrac{dq_{av}}{dt} = \dfrac{\omega^2 \delta v^2}{2v_0^2} \zeta .
\end{equation}

\subsection{Heat production in a generic regime}\label{sezioneVC}

In an arbitrary regime (with $\tau_H \gg \tau_m$) we need to perturb the telegraph-type equation 
\eqref{zazza}, obtaining
\begin{equation}
(\delta\indices{^A _B} - i \omega \tau\indices{^A _B})\delta \mathbb{A}^B = -i \omega  \dfrac{\delta v}{v_0} k_0^A,
\end{equation}
where we have omitted the subscript $0$ in $\tau\indices{^A _B}$. 
We define the matrix $\mathcal{M}\indices{^A _B}$ to be the inverse of $\delta\indices{^A _B} - i \omega \tau\indices{^A _B}$, which is proven to always exist in appendix \ref{TIOTTM}, so we have
\begin{equation}
\delta \mathbb{A}^A = -i \omega \dfrac{\delta v}{v_0} \mathcal{M}\indices{^A _B} k_0^B.
\end{equation}
Plugging this into \eqref{dqdt} we find
\begin{equation}
\dfrac{dq_{av}}{dt} = \dfrac{\omega^2 \delta v^2}{2 v_0^2} \Xi^0_{AB} \mathcal{M}\indices{^A _C} (\mathcal{M}\indices{^B _D})^* k_0^C k_0^D.
\end{equation}
By comparison with the parabolic limit it is natural to define an effective frequency-dependent bulk viscosity coefficient
\begin{equation}\label{bhuiop}
\zeta_{\textrm{eff}} = \Xi^0_{AB} \mathcal{M}\indices{^A _C} (\mathcal{M}\indices{^B _D})^* k_0^C k_0^D .
\end{equation}
Remembering that $\tau\indices{^A _B}$ is of the order of $\tau_M$, we obtain that in the limit $\omega \tau_M \rightarrow 0$ (which corresponds to $\tau_M \ll \tau_H$) $\mathcal{M}\indices{^A _B} \rightarrow \delta\indices{^A _B}$ and therefore $\zeta_{\textrm{eff}} \rightarrow \zeta$ and we recover the results given in subsection \ref{havana}. On the other hand, in the limit $\omega \tau_M \rightarrow +\infty$ (which corresponds to $\tau_M \gg \tau_H$) $\zeta_{\textrm{eff}}$ scales as $\omega^{-2}$. 
The physical explanation for this is that we are in the frozen limit when the oscillations are that fast. In fact, perturbing equation \eqref{strongG},
\begin{equation}
\delta x_A = \dfrac{i\Xi_{AB}^0 \delta \mathbb{A}^B}{n_0 \omega} = \delta v \, \Xi_{AB}^0 \mathcal{M}\indices{^B _C} k_0^C ,
\end{equation}
which goes to zero as $\omega \tau_M \rightarrow +\infty$. Since the production of entropy depends only on the average displacement of the fractions from equilibrium, which in this limit is determined only by the amplitude of the oscillation and not by its frequency, we have that $dq_{av}/dt$ must approach a constant value for large $\omega$, giving the condition $\omega^2 \zeta_{\textrm{eff}} \approx const$.

Equation \eqref{bhuiop} is the generalization to arbitrary $l$ (and arbitrary microscopic origin) of equation (10) of \citet{Sawyer_Bulk1989}. Note that the dependence on the frequency of $\zeta_{\textrm{eff}}$ does not disappear in the Newtonian limit, but constitutes the general thermodynamic explanation of the dependence on the frequency of the bulk viscosity noted by \citet{landau6} and \citet{Meador1996}.

\section{Bulk viscosity in neutron stars}\label{InNS}


Among the possible applications of our theory, a relevant one would be the study of the chemically induced bulk viscosity in neutron stars. Despite the fact that our formalism only provides the general form of the hydrodynamic equations, it can be adapted to include any kind of nuclear reaction which contributes to the bulk viscosity. In general, one should select the relevant chemical species (the number of chemical species will coincide with the number $l$), provide an equation of state (valid also out of chemical equilibrium) and give a formula for the reaction matrix $\Xi_{AB}$ (to do this one needs to expand the nuclear-reaction rates in the affinities).  Then, our machinery can be employed to compute the coefficients $\tau\indices{^A _B}$ and $k^A$ of the hyperbolic near-equilibrium model (see subsection \ref{expBuba!}) or, directly, the bulk viscosity $\zeta$. To give a practical example of how this procedure works, we will focus, here, on a minimal model in which dissipation is the product of beta-reactions.

We take a  two-component model, with number densities $n_p$ of protons and $n_n$ of neutrons. We require the fluid to be electrically neutral, so the density of electrons is not an independent degree of freedom. The equation of state is $\mathcal{U}= \mathcal{U}(s,n_p,n_n)$ whose differential is
\begin{equation}\label{xsqwe}
d\mathcal{U} =\Theta ds + \mu_n dn_n + \mu_p dn_p .
\end{equation}
Neglecting superfluidity and heat flux,   all the components are comoving with the entropy:
\begin{equation}
s^\nu = s u^\nu \spc n_p^\nu = n_p u^\nu  \spc  n_n^\nu = n_n u^\nu .
\end{equation}
As a result of $\beta$ reactions a particle of type $p$ can be converted into a particle of type $n$ and vice-versa, but the current
\begin{equation}
n^\nu = n_p^\nu +n_n^\nu
\end{equation}
is conserved. The differential \eqref{xsqwe} can be rewritten in the form
\begin{equation}
d\mathcal{U} =\Theta ds + \mu_n dn - \mathbb{A} dn_p ,
\end{equation}
where
\begin{equation}
\mathbb{A} = \mu_n - \mu_p 
\end{equation}
is the affinity of the reaction. Calling the fraction of $p$ particles $x_p = n_p/n$, the differential of the energy per-particle is
\begin{equation}
d\tilde{\mathcal{U}} = \Theta dx_s -\Psi dv - \mathbb{A} dx_p ,
\end{equation}
which is presented in the form \eqref{EOS}. We now rewrite the differential $d\tilde{\mathcal{U}}$ performing a different ``chemical choice'': the differential \eqref{xsqwe} could be equivalently given in the form
\begin{equation}
d\mathcal{U} =\Theta ds + \mu_p dn - \mathbb{B} dn_n ,
\end{equation}
with $\mathbb{B} = \mu_p - \mu_n = -\mathbb{A}$. Then, introducing the fraction of free neutrons $x_n = n_n/n = 1-x_p$, we would arrive at 
\begin{equation}
d\tilde{\mathcal{U}} = \Theta dx_s -\Psi dv - \mathbb{B} dx_n .
\end{equation}
This gauge fixing preserves $\Theta$ and $\Psi$, but not the chemical potential associated to the baryon current, which in the first case is $\mu_n$, in the second case is $\mu_p$. 

The equation of evolution of the dynamical fractions is
\begin{equation}\label{vnguir}
\dot{x}_p =\dfrac{1}{n} \Xi \mathbb{A}.
\end{equation}
In the limit of small affinities the telegraph-type equation reads
\begin{equation}
\tau_M \dot{\mathbb{A}} + \mathbb{A} = k \nabla_\nu u^\nu ,
\end{equation}
where, since the relaxation time matrix has only one element, we identified it with $\tau_M$ itself:
\begin{equation}\label{tauMMMMMMMMMMMMMMMMMMMMMM}
\tau_M = -\dfrac{n}{\Xi} \dfrac{\partial x_p}{\partial \mathbb{A}} \bigg|_{\mathbb{A}=0,v,x_s} .
\end{equation}
The quantity $k$ is
\begin{equation}
k = \dfrac{1}{\Xi} \dfrac{\partial x_p^{eq}}{\partial v} \bigg|_{x_s} .
\end{equation}

We can finally compute the formula for the effective bulk viscosity \eqref{bhuiop} by considering that in the case $l=2$ the matrix $\mathcal{M}\indices{^A _B}$ reduces to a single coefficient
\begin{equation}
\mathcal{M} = \dfrac{1}{1-i\omega \tau_M} .
\end{equation}
We immediately find that
\begin{equation}\label{sawyerrr}
\zeta_{\textrm{eff}} = \dfrac{\zeta}{1 + \omega^2 \tau_M^2} ,
\end{equation}
which is in accordance with \citet{Sawyer_Bulk1989}.

\section{Recovering Israel-Stewart}\label{recovering IS}

In this section we show how, starting from the general theory for $l=2$ (and only in this case)\footnote{
    For this reason we may say that, if for $l=0$ we have the barotropic perfect fluid, for $l=1$ the perfect fluid, then for $l=2$ we have the Israel-Stewart bulk-viscous fluid (we remind that $l$ gives the number of variables appearing in the equation of state apart from the baryon density).
}, one can recover the description for bulk viscosity of \citet{Israel_Stewart_1979}.

\subsection{Expanding the entropy}

In section \ref{eobv!!!} we have expanded the thermodynamic potential $\tilde{\mathcal{G}}$ for small affinities obtaining (in the first order limit) a hyperbolic version of relativistic Navier-Stokes with bulk viscosity. 
However, since the approach presented in sections \ref{TOOOEF} and \ref{DHOLIF} is completely general, we expect it  to reproduce also the Israel-Stewart description of bulk viscosity, taking the second order of the theory, and imposing $l=2$.

Let us introduce, for notational convenience, the quantities
\begin{equation}\label{def...iciente}
\beta := \dfrac{1}{\Theta}  \spc  \psi := \dfrac{\Psi}{\Theta}  \spc  a := \dfrac{\mathbb{A}}{\Theta} .
\end{equation}
Equation \eqref{entrppia} can be rewritten in the form (for $l=2$)
\begin{equation}\label{entrppia111}
dx_s = \beta \, d\tilde{\mathcal{U}} + \psi \, d v +  a \, dx .
\end{equation}
We introduce the new state variable 
\begin{equation}\label{legg333}
y_s = x_s -ax ,
\end{equation}
whose differential is
\begin{equation}\label{54321}
dy_s = \beta \, d\tilde{\mathcal{U}} + \psi \, d v -  x \, da .
\end{equation}
Analogously to what we did in section \ref{eobv!!!}, we expand $y_s$ to the second order in $a$:
\begin{equation}\label{9876}
y_s = x_s^{eq} - x^{eq} a +\dfrac{1}{2} y_s'' a^2 , 
\end{equation}
where $x_s^{eq}$, $x^{eq}$ and $y_s''$ are only functions of $\tilde{\mathcal{U}}$ and $v$. In the above formula we have used the fact that
\begin{equation}
x_s^{eq} = x_s|_{a=0} = y_s|_{a=0}
\end{equation}
and
\begin{equation}
x^{eq} = x|_{a=0} = -\dfrac{\partial y_s}{\partial a} \bigg|_{a=0} .
\end{equation}
Now we introduce the quantities
\begin{equation}
\beta^{eq} := \dfrac{\partial x_s^{eq}}{\partial \tilde{\mathcal{U}}}  \spc  \psi^{eq} := \dfrac{\partial x_s^{eq}}{\partial v} .
\end{equation}
Note that in section \ref{eobv!!!} the equilibrium quantities were defined as state variables of an hypothetical fluid in equilibrium with the same density and entropy per-particle, while now the hypothetical equilibrium configuration has the same density and energy per-particle. 
Therefore, at the second order $\beta^{eq} \neq \Theta_{eq}^{-1}$ and $\psi^{eq} \neq P/ \Theta_{eq}$, where $\Theta_{eq}$ and $P$ have been introduced in equation \eqref{zsefvgy}. 
Equation \eqref{polkj}, however, can be easily used to prove that they coincide at the first order. Comparing \eqref{9876} with \eqref{54321} we find that
\begin{equation}\label{machesbatti}
\begin{split}
& \beta = \beta^{eq} - a \dfrac{\partial x^{eq}}{\partial \tilde{\mathcal{U}}} + \dfrac{a^2}{2} \dfrac{\partial y_s''}{\partial \tilde{\mathcal{U}}} \\
& \psi = \psi^{eq} - a \dfrac{\partial x^{eq}}{\partial v} + \dfrac{a^2}{2} \dfrac{\partial y_s''}{\partial v} \\
& x = x^{eq} - y_s'' a . \\
\end{split}
\end{equation}
Now we can reverse the Legendre transformation \eqref{legg333} and get
\begin{equation}\label{polkj2}
x_s = x_s^{eq} - \dfrac{1}{2} y_s'' a^2 .
\end{equation}
Note that, since $a=0$ defines the maximum of the entropy, we immediately have
\begin{equation}
y_s'' >0 .
\end{equation}
Defining $s_{eq}:= nx_s^{eq}$ and using the definition of $a$ \eqref{def...iciente} we find
\begin{equation}\label{timeodanaosetdonaferentes}
s = s_{eq} - \dfrac{n y_s''}{2 \Theta^2} \mathbb{A}^2 .
\end{equation}
Now, to compare equation \eqref{timeodanaosetdonaferentes} with the expansion of Israel and Stewart \citep{andersson2007review}, we need to use the quantity $\Pi$ as a free variable in the equation of state of the entropy. 
In our approach, $\Pi$ is not a fundamental thermodynamic quantity, but it is the first order correction to $\Psi$ when the matter element goes out of equilibrium. Now, there is the complication that the reference equilibrium state which is considered in section \ref{eobv!!!} is different from the one assumed in this section. Luckily, this does not produce any confusion in the definition of $\Pi$, because this distinction emerges only at the second order. Therefore, we can still employ equation \eqref{qwerty}, imposing $l=2$, and find
 \begin{equation}\label{viscstrwiw}
\Pi = -\mathbb{A} \dfrac{\partial x^{eq}}{\partial v} \bigg|_{x_s},
\end{equation}
which, plugged into \eqref{timeodanaosetdonaferentes}, gives
\begin{equation}
s = s_{eq} - \dfrac{\chi \Pi^2}{2 \Theta},
\end{equation}
with
\begin{equation}\label{matuchicazzosei}
\chi = \dfrac{n y_s''}{\Theta} 
\bigg( \dfrac{\partial x^{eq}}{\partial v} \bigg|_{x_s} \bigg)^{-2} \geq 0 .
\end{equation}
This is the second order expansion of the entropy in terms of $\Pi$ proposed in the Israel-Stewart formulation, see \cite{andersson2007review}. We remark that they denote our $\chi$ with the symbol $\beta_0$, but we have changed notation to avoid confusion with the inverse temperature.

The key step in the construction of a bridge between the formalisms is to find an algebraic relationship between the parameter $\chi$ introduced in Israel-Stewart theory, and the coefficients presented in section \ref{eobv!!!}. This can be done deriving the third equation of \eqref{machesbatti} with respect to $\mathbb{A}$ at constant $v$ and $x_s$, obtaining
\begin{equation}\label{krabnebula}
\dfrac{\partial x}{\partial \mathbb{A}} \bigg|_{v,x_s} = - \dfrac{y_s''}{\Theta}.
\end{equation}
Plugging this result into \eqref{matuchicazzosei}, with the aid of \eqref{TTau} and \eqref{zetta}, we obtain the formula
\begin{equation}\label{chichichcichci}
\tau_M = \chi \zeta .
\end{equation}

\subsection{Telegraph-type equation for the viscous stress}
 
Equation \eqref{zazza} for the case $l=2$ reads
\begin{equation}
\tau_M \dot{ \mathbb{A}} + \mathbb{A} = k \nabla_\nu u^\nu .
\end{equation}
Using the definition for the transport coefficients $k$ and $\zeta$, equations \eqref{kappa} and \eqref{zetta}, and the formula \eqref{viscstrwiw}, we get
\begin{equation}
\Pi = -\zeta \nabla_\nu u^\nu + \tau_M \dot{\mathbb{A}} \dfrac{\partial x^{eq}}{\partial v} \bigg|_{x_s},
\end{equation}
which can be rewritten in the form
\begin{equation}
\Pi = -\zeta \nabla_\nu u^\nu - \tau_M \dot{\Pi}  - \tau_M \mathbb{A} u^\nu \partial_\nu \bigg( \dfrac{\partial x^{eq}}{\partial v} \bigg|_{x_s} \bigg).
\end{equation}
Neglecting higher order terms, we use \eqref{chichichcichci} to rewrite the above equation in the form
\begin{equation}
\Pi = -\zeta ( \nabla_\nu u^\nu + \chi \dot{\Pi} )  - \mathbb{A} \zeta \nabla_\nu u^\nu   \dfrac{\partial^2 x^{eq}}{\partial v^2} \bigg|_{x_s} \chi v  .
\end{equation}
The last term in the right-hand side is proportional to the product between $\mathbb{A}$ and $\zeta \nabla_\nu u^\nu$, therefore it is a higher order with respect to the other terms and we can neglect it. In the end we obtain the equation
\begin{equation}
\Pi = -\zeta ( \nabla_\nu u^\nu + \chi \dot{\Pi} ),
\end{equation}
which completes the bridge between our formulation and the one of \citet{Israel_Stewart_1979}. 
We remark that it has been shown in \cite{Causality_bulk} that this model for bulk viscosity is causal.

\subsection{Israel-Stewart modelling of neutron star bulk viscosity}

To complete our discussion let us show how the coefficients of the Israel-Stewart expansion above can be computed from a multifluid approach for the case of neutron star matter presented in section \ref{InNS}.

The viscous stress $\Pi$, which in \cite{Israel_Stewart_1979} is treated as a fundamental variable, can be written in terms of quantities appearing in the two-fluid model as
\begin{equation}
\Pi=(\mu_p-\mu_n)\dfrac{\partial x_p^{eq}}{\partial v} \bigg|_{x_s} ,
\end{equation}
see equation \eqref{qwerty}. The transport coefficient $\zeta$ is given in general by the formula \eqref{zetta}, which in our case reduces to
\begin{equation}\label{wswdefrgr}
\zeta = \dfrac{1}{\Xi} \bigg( \dfrac{\partial x_p^{eq}}{\partial v} \bigg|_{x_s} \bigg)^2,
\end{equation}
where $\Xi$ has been introduced in \eqref{vnguir}.\footnote{
    In the literature \citep{Sawyer_Bulk1989,Gusakov2007}, $\Xi$ is usually denoted by $\lambda$ and equation \eqref{vnguir} is expressed in the different notation $\Gamma = \lambda \, \delta \mu$. } 
In the literature it is possible to find this formula written in terms of a different set of thermodynamic quantities. Let us define the neutron excess as $\alpha = x_n-x_p$ and consider the equation
\begin{equation}
\mathbb{A} = \mathbb{A}(n,\alpha,x_s).
\end{equation}
If we derive it along the curve
\begin{equation}
\alpha=\alpha^{eq} \spc x_s = const ,
\end{equation}
we find the thermodynamic relation
\begin{equation}\label{eiufnioenovmv9uo}
\dfrac{\partial x_p^{eq}}{\partial v} \bigg|_{x_s} = -\dfrac{n^2 \partial_n \mathbb{A}|_{\alpha,x_s}}{2\partial_\alpha \mathbb{A}|_{n,x_s}}, 
\end{equation}
where from now on, in this subsection, everything is computed in equilibrium, i.e. for $\alpha= \alpha^{eq}$. Plugging \eqref{eiufnioenovmv9uo} into \eqref{wswdefrgr}, we obtain
\begin{equation}\label{zzzetestx}
\zeta = \dfrac{n^4 ( \partial_n \mathbb{A}|_{\alpha,x_s})^2}{4\Xi( \partial_\alpha \mathbb{A}|_{n,x_s})^2}     . 
\end{equation}
The times-scale $\tau_M$ is given in \eqref{tauMMMMMMMMMMMMMMMMMMMMMM} and can be rewritten in the form
\begin{equation}\label{mmmmmMmmmmMmmm}
\tau_M = \dfrac{n}{2\Xi \partial_\alpha \mathbb{A}|_{n,x_s}}.
\end{equation}
If we plug \eqref{zzzetestx} and \eqref{mmmmmMmmmmMmmm} into \eqref{sawyerrr} we get the formula for $\zeta_{\textrm{eff}}$ that you can find in \cite{Sawyer_Bulk1989} and \cite{Haensel_Bulk_Urca}.

Finally, we can use equation \eqref{chichichcichci} to compute the coefficient $\chi$ and get
\begin{equation}
\chi = \dfrac{2 \partial_\alpha \mathbb{A}|_{n,x_s}}{n^3 (\partial_n \mathbb{A}|_{\alpha,x_s})^2}.
\end{equation}

\section{Connection to microphysics: ideal gases}\label{abcdefghi}

We are now ready to connect the hydrodynamic description developed in sections \ref{TOOOEF} and \ref{DHOLIF} directly with a kinetic theory of ideal simple (i.e. without internal degrees of freedom other from the spin) gases. In particular we will prove directly from our formalism that the second viscosity must vanish in the non-relativistic and in the ultra-relativistic limit (c.f. \cite{landau10}). We will, then, present the equation of state of the gas, extended to quasi-equilibrium states, for the intermediate case, in a minimal model with $l=2$.

\subsection{Elements of kinetic theory}\label{ottopuntouno}

We need to specialise the analysis presented in section \ref{TOOOEF} to a gas in which interactions are given only by instantaneous collisions, in the limit of small cross sections. The single particle Hamiltonian is assumed invariant under spin flip, so there is full degeneracy in the spin. 
For clarity we  present the derivation step by step. 

Consider a homogeneous portion of the gas in a box with reflecting walls. 
The state of this gas is described with a distribution function $f=f(\vect{x},\vect{p})$, which is the number of particles in per unit single-particle phase space volume,
\begin{equation}
f = \dfrac{dN}{d_3 x \, d_3 p}.
\end{equation}
Since we work only in the frame identified by the box, we do not need to study the behaviour of $f$ under Lorentz transformations. Because of homogeneity and isotropy, $f$ depends only on the modulus of $\vect{p}$ and describes a uniform local property of the matter elements. 
Note that if $f$ were not isotropic the interaction of the particles with the walls could alter the value of $f$ with time. However, since a collision sends a component $p_j \rightarrow -p_j$, an isotropic $f$ is on average unaltered by this process.

We can use $f$ to compute the particle density
\begin{equation}\label{densità}
n = \int f \, d_3 p ,
\end{equation}
the energy density
\begin{equation}\label{energgia}
\mathcal{U} = \int \epsilon \, f \, d_3 p 
\end{equation}
and the isotropic stress
\begin{equation}\label{pressszione}
\Psi =\dfrac{1}{3} \int  p_j v^j \, f \, d_3 p ,
\end{equation}
where we have that $\epsilon = \epsilon (\vect{p})$ is the single-particle energy and
\begin{equation}\label{VvvV}
v^j = \dfrac{\partial \epsilon}{\partial p_j}.
\end{equation}
It is also possible to compute the entropy per unit volume
\begin{equation}\label{SssS}
s = \int \sigma \, f \, d_3 p,
\end{equation}
where
\begin{equation}\label{sigma}
\sigma = -\ln \bigg( \dfrac{f h_p^3}{g} \bigg) + \bigg( 1- \dfrac{g\iota}{ f h_p^3} \bigg) \ln \bigg( 1- \dfrac{\iota fh_p^3}{g} \bigg).
\end{equation}
$h_p$ is the Planck constant, $g$ is a possible degeneracy of spin and $\iota$ is a coefficient which is equal to $-1$ for Bosons, to $+1$ for Fermions and to $0$ in the classic limit. 

Now we need to identify the two time-scales $\tau_m$ and $\tau_M$. We define $\tau_m$ to be the typical time necessary for a particle to cross the box and $\tau_M$ to be inverse of the frequency of the binary collisions (which are assumed to be the dominant relaxation process). One may raise the criticism that the hydrodynamic description can exist only if the collision frequency is larger than the characteristic hydrodynamic frequencies, implying $\tau_H \gg \tau_M$. This would lead to the claim that a hyperbolic hydrodynamic formulation for bulk viscosity in this case is not guaranteed to exist, but that only its parabolic and perfect-fluid limits are possible, see subsection \ref{timescales!!!!}. This serious issue is connected with the fact that $l$ may a priori be infinite, as will be explained in more detail in subsection \ref{explofdegree}. In this section, however, we will assume that, for all practical applications, it is possible to approximate the system with a finite $l$ model. 

The assumption $\tau_m \ll \tau_M$ means that we are assuming that particles collide with the walls infinitely more often than with each other. During an expansion occurring in the time-scale $\tau_{fr}$ introduced in section \ref{TOOOEF} the particles do not have time to interact, but slam against the walls with an infinite frequency with respect to the rate of change of the position of the walls. In appendix \ref{sec:adexp} we show that if we parametrise the expansion with $\lambda$, such that $v \rightarrow \lambda^3 v$, then we have that in this process $f \rightarrow f_\lambda$, with
\begin{equation}\label{adexpsion}
f_\lambda (\vect{p}) = f(\lambda \vect{p}).
\end{equation}
We also prove that this transformation satisfies the conditions
\begin{equation}
 \dfrac{dx_s}{d\lambda} =0 \spc  \spc  \dfrac{d \tilde{\mathcal{U}}}{d\lambda} = -\Psi \dfrac{d v}{d \lambda}.
\end{equation}
This is coherent with equations \eqref{itregrassi}, \eqref{KkK} and \eqref{grandpotepress}.

\subsection{Constructing the manifold of quasi-equilibrium states}

Now we need to introduce the manifold $\mathcal{Z}$ of quasi-equilibrium states. If  $f$ is a generic isotropic function, then the number $l$ goes to infinity, because $\mathcal{Z}$ should coincide with the Banach space $\mathfrak{B}$ of the isotropic functions, which has infinite dimensions. 
To produce a conceivable hydrodynamic description directly from the kinetic theory one has to assume that only a limited subset of $\mathfrak{B}$ is sufficient to describe the quasi-equilibrium states the fluid will explore. Here we show the general strategy to obtain this manifold.

First of all we note that the two-dimensional manifold of the local equilibrium macrostates has to be a submanifold of $\mathcal{Z}$. 
For a gas, the equilibrium distribution, i.e. the one which maximizes the entropy \eqref{SssS}, fixed the density of particles and the energy density, and which will, therefore, satisfy \eqref{equilllll} whichever set of $\alpha_A$ we choose, is
\begin{equation}
f_{\beta,\mu}(\vect{p}) =\dfrac{g}{h_p^3} \dfrac{1}{e^{\beta (\epsilon(\vect{p})-\mu)} + \iota},
\end{equation}
where $\beta$ is the inverse of the temperature and $\mu$ is the chemical potential. So we need to impose $f_{\beta,\mu} \in \mathcal{Z}$.

Secondly, we note that the adiabatic expansion \eqref{adexpsion} must describe a group of transformations on $\mathcal{Z}$ and must not send a point of $\mathcal{Z}$ out of it. In mathematical terms, we may say that $\mathcal{Z}$ must be invariant under the group of the adiabatic expansions, which send $f_{\beta,\mu}$ into
\begin{equation}\label{lambdato}
f_{\beta,\mu,\lambda}(\vect{p}) =\dfrac{g}{h_p^3} \dfrac{1}{e^{\beta (\epsilon(\vect{\lambda p})-\mu)} + \iota}.
\end{equation}
So the set of the $f_{\beta,\mu,\lambda}$ must be a submanifold of $\mathcal{Z}$.  Note that for $\lambda \neq 1$ it is not necessarily true that $\beta$ and $\mu$ can be interpreted as inverse temperature and chemical potential. This is connected to the chemical gauge freedom discussed in subsection \ref{gauge}. 

Finally, the manifold should be extended considering the collision processes. To understand how, imagine a homogeneous portion of a gas at rest in a box, prepared in an arbitrary out-of-equilibrium state. The collisions will tend to drive $f$ towards equilibrium following a curve $f(t)$. If we want to be able to fully describe this curve in the framework of our hydrodynamic description we need $f(t)$ to be a map
\begin{equation}
f: \mathbb{R} \longrightarrow \mathcal{Z}\subset \mathfrak{B}.
\end{equation}  
On the other hand, the curve $f(t)$ is governed by an equation of the form
\begin{equation}
\dfrac{df}{dt} = \dot{f}_{coll}[f],
\end{equation}
where the right-hand side is a collision functional. Therefore we need $\mathcal{Z}$ to be an invariant set of the flux generated by $\dot{f}_{coll}$. 

In practice, however, finding the submanifold of $\mathfrak{B}$ which contains the equilibrium states and is invariant under both the adiabatic expansion and the flux generated by $\dot{f}_{coll}$ can be hard, because the two generators do not commute, 
\begin{equation}\label{ghjlòo}
\bigg[ \dfrac{d}{d\lambda} , \dot{f}_{coll} \bigg] \neq 0,
\end{equation}
where $d/d\lambda$ and $\dot{f}_{coll}$ are seen as vector fields tangent to $\mathfrak{B}$. To prove the inequality \eqref{ghjlòo}, imagine to start in an equilibrium state: if we make a relaxation process followed by an adiabatic expansion, if the bulk viscosity is not zero the result is an out-of-equilibrium state. On the other hand, if we invert the order of the transformations and assume that the relaxation process is sufficiently long, we may end up in thermodynamic equilibrium.

In general, the dimension of $\mathcal{Z}$ may be arbitrarily high (and we will explain the consequences of this problem in subsection \ref{explofdegree}) so it can be convenient to make an hierarchy of approximated theories with increasing $l$, starting with the minimal $l=2$ manifold of $f_{\beta,\mu,\lambda}$ and expanding the set of included functions gradually, to increase the precision. 

In this work we will restrict ourselves to the  $l=2$ model, which will allow us to have a direct comparison with the microphysical calculations of \citet{Israel_Stewart_1979}.


\subsection{Proving that the second viscosity of non-relativistic and ultra-relativistic ideal gases vanishes}\label{PTTSVONRAURGV}

A non-relativistic gas is described through the dispersion law
\begin{equation}
\epsilon (\vect{p}) = m + \dfrac{\vect{p}^2}{2 m} .
\end{equation}
Plugging this condition inside \eqref{lambdato} we find that
\begin{equation}\label{nonaumenta}
f_{\beta,\mu,\lambda} = f_{\beta' , \mu'}
\end{equation}
with
\begin{equation}\label{bmuprim1}
\beta' = \beta \lambda^2  \spc  \mu' = m + \dfrac{\mu -m}{\lambda^2}.
\end{equation}
An ultra-relativistic gas, on the other hand, is obtained imposing that the single-particle energy has the form
\begin{equation}
\epsilon (\vect{p}) = |\vect{p}|.
\end{equation}
This, plugged in equation \eqref{lambdato}, gives again a relation of the type \eqref{nonaumenta}, with the transformation
\begin{equation}\label{bmuprim2}
\beta' = \lambda \beta  \spc  \mu' = \dfrac{\mu}{\lambda}.
\end{equation}
Combining the first equations of \eqref{bmuprim1}  and \eqref{bmuprim2} with the fact that $v'=\lambda^3 v$ we have verified that they follow the equilibrium adiabatic curves
\begin{equation}
\Theta' v'^{\Gamma -1} = \Theta v^{\Gamma-1} ,
\end{equation}
where $\Gamma = 5/3$ for the non-relativistic gas and $4/3$ for the ultra-relativistic gas. So we have proved that in both cases the adiabatic curves, generated in the fast expansion, which start in an equilibrium state, remain in the surface of the equilibrium states. This implies that if a fluid element is prepared in a local thermodynamic equilibrium state, then the expansions and contractions it will incur will not be able to drive it out of equilibrium, independently from the speed of the expansion/contraction, provided that $\tau_H \gg \tau_m$, proving that in these two cases the bulk viscosity vanishes, as explained also in \cite{landau10} and \cite{Tisza_Bulk}. An alternative proof of this fact, based on geometrical arguments, is given in appendix \ref{staiqui}.

\subsection{The equation of state for a relativistic diluted massive gas out of equilibrium}\label{TEOSFARDMGOOE}

Let us impose $\iota =0$ and
\begin{equation}
\epsilon(\vect{p}) = \sqrt{m^2 + \vect{p}^2}.
\end{equation}
Then the distribution in a generic state of $\mathcal{Z}$ is
\begin{equation}\label{lanostraflaNOOOOSTRA}
f_{\beta,\mu,\lambda} =\dfrac{g}{h_p^3} exp\bigg[ \alpha - \zeta_c \sqrt{1 + \dfrac{\lambda^2 \vect{p}^2}{m^2} }  \bigg],
\end{equation}
where we have introduced
\begin{equation}\label{interpretoinuoviarrivati}
\alpha = \beta \mu  \spc  \zeta_c = \beta m .
\end{equation}
In the case $\lambda = 1$, when the fluid is in thermal equilibrium, $\alpha$ is the fugacity and $\zeta_c$ is the coldness \cite{rezzolla_book}, but out of equilibrium this interpretation is lost because $\beta$ and $\mu$ are no longer the inverse temperature and chemical potential of the fluid. The variables $(\lambda,\alpha,\zeta_c)$ define a chart on $\mathcal{Z}$ and the curves generated by $W_{fr}$ are curves with constant $\alpha$ and $\zeta_c$. This means that both $\alpha$ and $\zeta_c$ are good candidates to become the variable we can use to parametrise the quasi-equilibrium states.

We define the factor
\begin{equation}
b := \dfrac{4\pi g m^3}{h_p^3}
\end{equation}
and the function
\begin{equation}
\phi(\zeta_c) :=\ln \int_0^{+\infty} \xi^2 e^{-\zeta_c \sqrt{1+\xi^2}}  \, d\xi,
\end{equation}
whose first and second derivative in $\zeta_c$ will be denoted respectively by $\phi'$ and $\phi''$.
Using equations \eqref{densità} and \eqref{SssS}, we find
\begin{equation}\label{ledueragazze}
v = \dfrac{\lambda^3}{b  e^{\alpha+\phi}}  \spc  x_s = -\alpha - \zeta_c \phi'.
\end{equation}
Note that $x_s$ does not depend on $\lambda$ and this is coherent with the fact that it is conserved along the curves generated by $W_{fr}$. We are, now, able to introduce the chart $(v,x_s,\zeta_c)$, which is in the form discussed in section \ref{GEOS}. We can invert the foregoing equations obtaining
\begin{equation}\label{coordintrasnfr}
\alpha = -x_s - \zeta_c \phi'  \spc  \lambda = \big( bv e^{\phi-\zeta_c \phi'-x_s} \big)^{1/3}.
\end{equation}  
The distribution $f$ is naturally given in terms of the parameters $\lambda$, $\alpha$ and $\zeta_c$, so it is natural to use these variables while performing an average to extract a thermodynamic variable. Then, making the coordinate transformation
\begin{equation}
(\lambda,\alpha,\zeta_c)  \quad \longmapsto \quad (v,x_s,\zeta_c)
\end{equation}
by means of the formulas \eqref{coordintrasnfr}, the quantity can be finally written as a function of multifluid-type variables. In particular this can be done to obtain the equation of state $\tilde{\mathcal{U}}(v,x_s,\zeta_c)$. In fact, recalling equation \eqref{energgia} and making use of the first equation of \eqref{ledueragazze}, we find 
\begin{equation}
\tilde{\mathcal{U}}(\lambda,\zeta_c) = m e^{-\phi}  \int_0^{+\infty} \xi^2 e^{-\zeta_c \sqrt{1+\xi^2}} \sqrt{1+ \dfrac{\xi^2}{\lambda^2} } \, d\xi .
\end{equation}
We remark that the fact that $\tilde{\mathcal{U}}$ does not depend on $\alpha$ is a useful result which holds only in the non-degenerate limit.
Making the change of variables we finally obtain the equation of state for the quasi-equilibrium gas:
\begin{equation}\label{equazdistato}
\begin{split}
& \tilde{\mathcal{U}}(v,x_s,\zeta_c) =  me^{-\phi} \\
 & \int_0^{+\infty} \xi^2 e^{-\zeta_c \sqrt{1+\xi^2}} \sqrt{1+ \xi^2 \bigg( \dfrac{e^{x_s -\phi+ \zeta_c \phi'} }{bv} \bigg)^{2/3} } \, d\xi .\\
 \end{split} 
\end{equation}

\begin{figure*}
\begin{center}
	\includegraphics[width=1.0\textwidth]{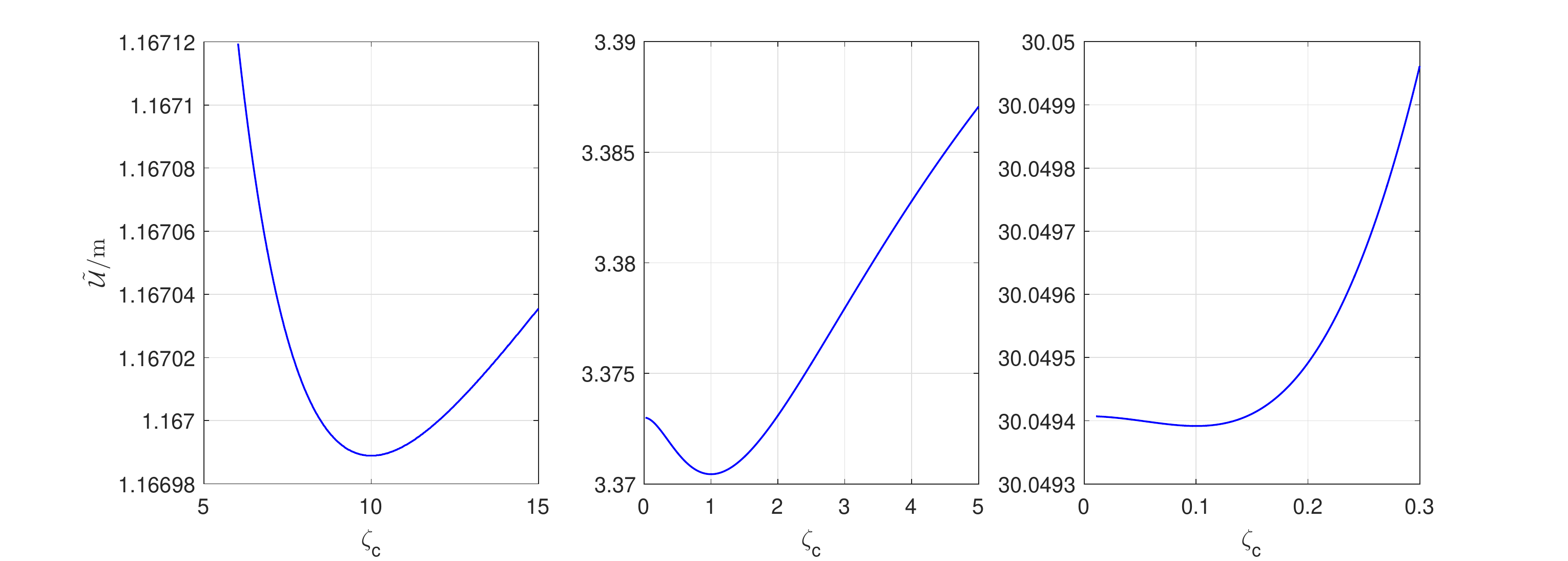}
	\caption{Plots of $\tilde{\mathcal{U}}/m$ as a function of $\zeta_c$, with $v$ and $x_s$ fixed and chosen in a way that: $\zeta_c^{eq}=10$ (first panel), $\zeta_c^{eq}=1$ (second panel), $\zeta_c^{eq}=0.1$ (third panel).}
	\label{fig:energy}
	\end{center}
\end{figure*}

In figure \ref{fig:energy} it is possible to see three plots representing $\tilde{\mathcal{U}}/m$ as a function of $\zeta_c$. $v$ and $x_s$ are held fixed and chosen is such a way that when $\lambda=1$, i.e. in thermal equilibrium, $\zeta_c$ takes the values respectively $10$ (first panel), $1$ (second panel) and $0.1$ (third panel). The energy has its absolute minimum in the equilibrium state, coherently with the results of section \ref{EFSDFETP}. 


%
%


\subsection{Pressure, temperature and affinity in the quasi-equilibrium states}\label{2TEOSFARDMGOOE}

To compute the derivatives of the energy per particle it is convenient to pass through the variable $\lambda$. For example, to compute the pressure, according to \eqref{EOS} and \eqref{grandpotepress}, we only need to calculate
\begin{equation}\label{xmdjruyt}
\Psi = - \dfrac{\partial \tilde{\mathcal{U}}}{\partial v} \bigg|_{x_s,\zeta_c} =  - \dfrac{\partial \tilde{\mathcal{U}}}{\partial \lambda} \bigg|_{\zeta_c} \dfrac{\partial \lambda}{\partial v} \bigg|_{x_s,\zeta_c} .
\end{equation}
Using \eqref{coordintrasnfr}, we see that
\begin{equation}\label{bvjfirut}
\dfrac{\partial \lambda}{\partial v} \bigg|_{x_s,\zeta_c} =  \dfrac{\lambda}{3v} ,
\end{equation}
and we get the formula
\begin{equation}
\Psi = \dfrac{m e^{-\phi}}{3v}  \int_0^{+\infty}  e^{-\zeta_c \sqrt{1+\xi^2}}   \dfrac{\xi^4/\lambda^2}{\sqrt{1+ \xi^2/\lambda^2  }} \, d\xi .
\end{equation}
It is possible to check with a little algebra that this coincides with \eqref{pressszione}, proving the consistency of the formulation.

The temperature, according to \eqref{EOS}, is
\begin{equation}
\Theta =  \dfrac{\partial \tilde{\mathcal{U}}}{\partial x_s} \bigg|_{v,\zeta_c} =   \dfrac{\partial \tilde{\mathcal{U}}}{\partial \lambda} \bigg|_{\zeta_c} \dfrac{\partial \lambda}{\partial x_s} \bigg|_{v,\zeta_c} .
\end{equation}
However, it is true that
\begin{equation}
\dfrac{\partial \lambda}{\partial x_s} \bigg|_{v,\zeta_c} = -\dfrac{\lambda}{3},
\end{equation}
therefore, combining with \eqref{xmdjruyt} and \eqref{bvjfirut} we find the equation
\begin{equation}
\Psi = n \Theta .
\label{idealgaslawww}
\end{equation}
Thus we have found that the ideal gas law, which must be verified in equilibrium, remains true also out of it, provided that we choose $\zeta_c$ to be the additional variable in the equation of state.

The variable $\mathbb{A}$ is calculated in appendix \ref{Aoadrg}. We also verify there that in equilibrium (when $\lambda=1$) it vanishes, in agreement with the fact that this is the condition of thermodynamic equilibrium. This is also in accordance with the plots of figure \ref{fig:energy}.

\subsection{The Israel-Stewart limit of the theory}\label{3TEOSFARDMGOOE}

The full equation of state can be expanded for small deviations from equilibrium. 
As  shown in section \ref{recovering IS}, this will lead us to Israel-Stewart theory for bulk viscosity. 

\begin{figure}
\begin{center}
	\includegraphics[width=0.5\textwidth]{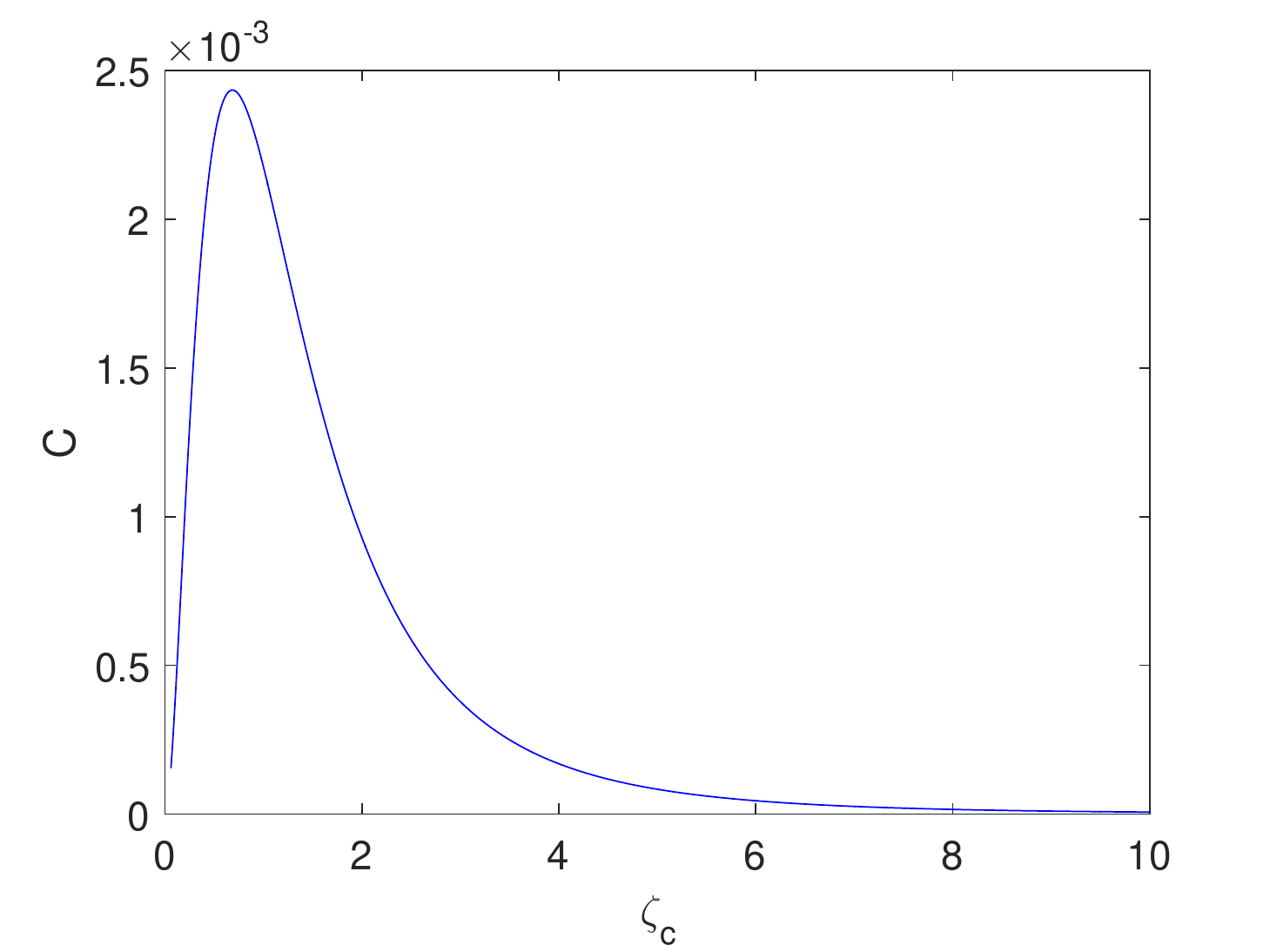}
	\caption{Plot of the susceptibility $\mathcal{C}$ as a function of $\zeta_c$.}
	\label{fig:susc}
	\end{center}
\end{figure}

Let us define the functions
\begin{equation}
J(\zeta_c):=\dfrac{e^{-\phi}}{3} \int_0^{+\infty} e^{-\zeta_c \sqrt{1+\xi^2}}   \dfrac{\xi^4}{(1+ \xi^2  )^{3/2}} \, d\xi
\end{equation}
and
\begin{equation}
G(\zeta_c) := \dfrac{K_3 (\zeta_c)}{K_2(\zeta_c)},
\end{equation}
where $K_\nu$ is the $\nu$-th modified Bessel function of the second type. Then it can be shown (see \cite{rezzolla_book} and references therein) that
\begin{equation}\label{conversion}
\begin{split}
& \phi = \ln \bigg( \dfrac{K_2}{\zeta_c} \bigg) \\
& \phi' = \dfrac{1}{\zeta_c} - G  \\  
& \phi''= -\dfrac{1}{\zeta_c^2} - G^2 + \dfrac{5G}{\zeta_c} + 1.
\end{split}
\end{equation}
In appendix \ref{TSDOTEITGRC} we prove that
\begin{equation}
\dfrac{\partial^2 \tilde{\mathcal{U}}}{\partial \zeta_c^2}  \bigg|_{v,x_s,\zeta_c=\zeta_c^{eq}} = \dfrac{m \zeta_c {\phi''}^2}{3} \bigg( 1+ \zeta_c J - \dfrac{3}{\zeta_c^2 \phi''} \bigg).
\end{equation} 
The function
\begin{equation}\label{ssssusssccceptToooooo!}
\mathcal{C} := \dfrac{1}{\tilde{\mathcal{U}}_{eq}} \dfrac{\partial^2 \tilde{\mathcal{U}}}{\partial \zeta_c^2} \bigg|_{v,x_s,\zeta_c=\zeta_c^{eq}}
\end{equation}
describes the susceptibility of the fluid to a displacement from equilibrium. In figure \ref{fig:susc} it is possible to see the behaviour of $\mathcal{C}$ as $\zeta_c$ varies. In the limits $\zeta_c \longrightarrow 0$ (ultra-relativistic limit) and $\zeta_c \longrightarrow + \infty$ (non-relativistic limit) it goes to zero, while it has its maximum around $\zeta_c =1$. This is a result of the fact that in the two opposite limits the system becomes degenerate in $\zeta_c$ and the equation of state depends only on two independent variables: $\tilde{\mathcal{U}}(v,x_s,\zeta_c)= \tilde{\mathcal{U}}(v,x_s)$, cf. with section \ref{PTTSVONRAURGV}. 

We can finally compute the thermodynamic coefficient $\chi$ introduced in equation \eqref{matuchicazzosei} and compare it with the prediction of \cite{Israel_Stewart_1979}.

Since equation \eqref{equazdistato} cannot be easily inverted to write the entropy as a function of $v$, $\tilde{\mathcal{U}}$ and $\zeta_c$, it is more convenient to recast \eqref{matuchicazzosei} in a form which involves derivatives of the energy per particle. Equation \eqref{krabnebula} can be easily used to prove that
\begin{equation}
\dfrac{y_s''}{\Theta} = \bigg( \dfrac{\partial^2 \tilde{\mathcal{U}}}{\partial \zeta_c^2} \bigg|_{v,x_s} \bigg)^{-1},
\end{equation}
where the second derivative is evaluated in equilibrium. This, plugged in \eqref{matuchicazzosei}, gives
\begin{equation}\label{berghione}
\chi = \dfrac{n}{\tilde{\mathcal{U}} \mathcal{C}} \bigg( \dfrac{\partial \zeta_c^{eq}}{\partial v} \bigg|_{x_s} \bigg)^{-2} .
\end{equation}
Since in equilibrium $\lambda=1$, taking the logarithm of the second equation of \eqref{coordintrasnfr} we find that $\zeta_c^{eq}(v,x_s)$ has to satisfy the condition
\begin{equation}
\ln b + \phi(\zeta_c^{eq}) + \ln v -x_s - \zeta_c^{eq} \phi'(\zeta_c^{eq}) =0.
\end{equation} 
Deriving with respect to $v$, keeping $x_s$ fixed, we obtain
\begin{equation}
\dfrac{\partial \zeta_c^{eq}}{\partial v}  \bigg|_{x_s} = \dfrac{1}{v \, \zeta_c^{eq} \phi''(\zeta_c^{eq})} .
\end{equation}
So we finally find
\begin{equation}
\chi = \dfrac{3}{P_{eq}} \bigg( 1+ \zeta_c J - \dfrac{3}{\zeta_c^2 \phi''} \bigg)^{-1} .
\end{equation}
This can be compared with the prediction of \citet{Israel_Stewart_1979} (based on the Grad 14-moment approximation)
\begin{equation}
\chi_{IS} = \dfrac{3}{P_{eq}} \, \dfrac{5 - 3 \gamma + 3(10-7\gamma)G/\zeta_c}{G^2 (3\gamma -5 +3\gamma/G \zeta_c)^2},
\end{equation}
where $\gamma$ is defined through the equation
\begin{equation}
\dfrac{\gamma}{\gamma -1} = \zeta_c^2 \bigg( 1 + \dfrac{5G}{\zeta_c} -G^2 \bigg).
\end{equation}

\begin{figure}
\subfigure{
\includegraphics[width=0.5\textwidth]{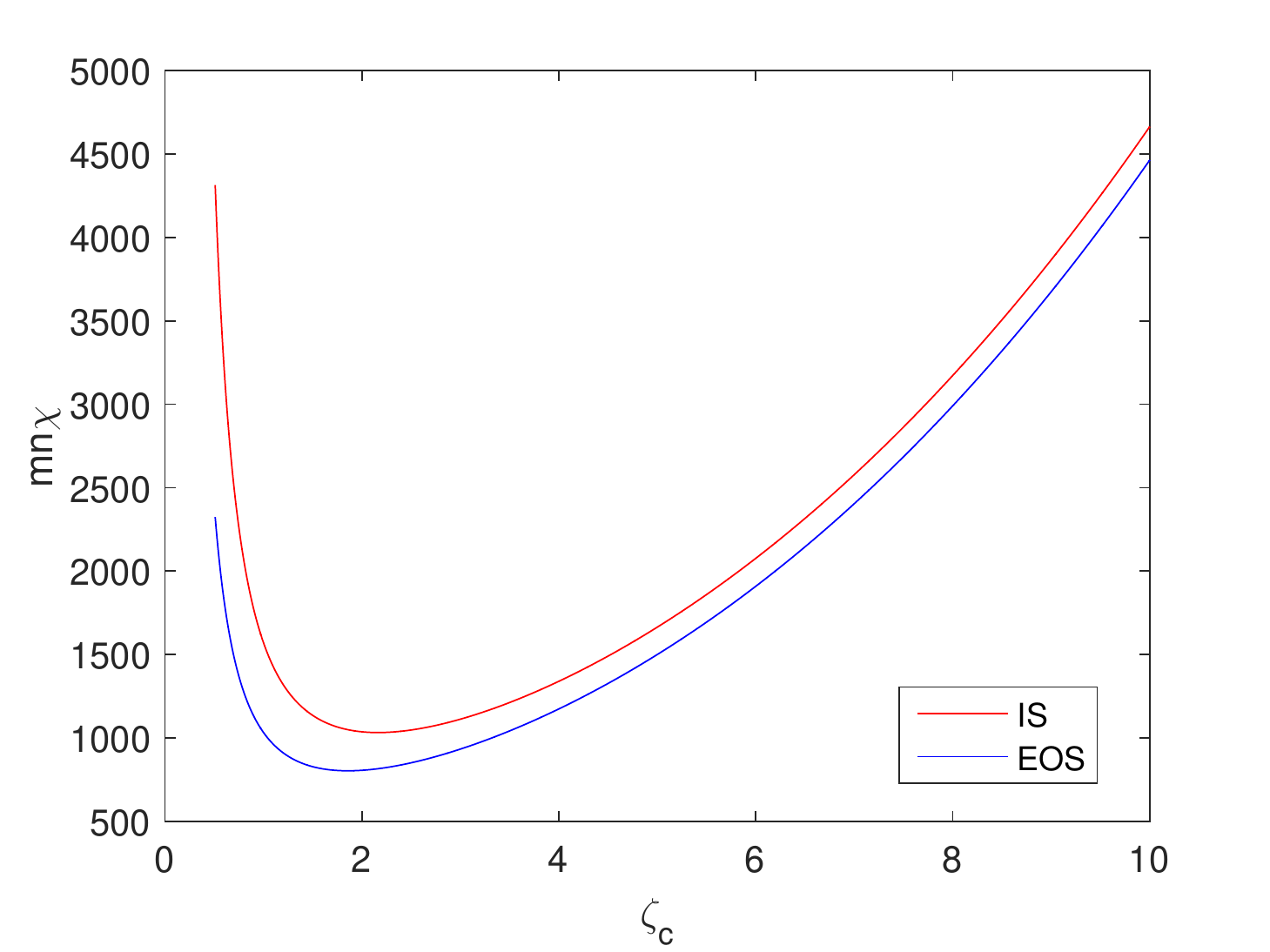} }
\subfigure{
\includegraphics[width=0.5\textwidth]{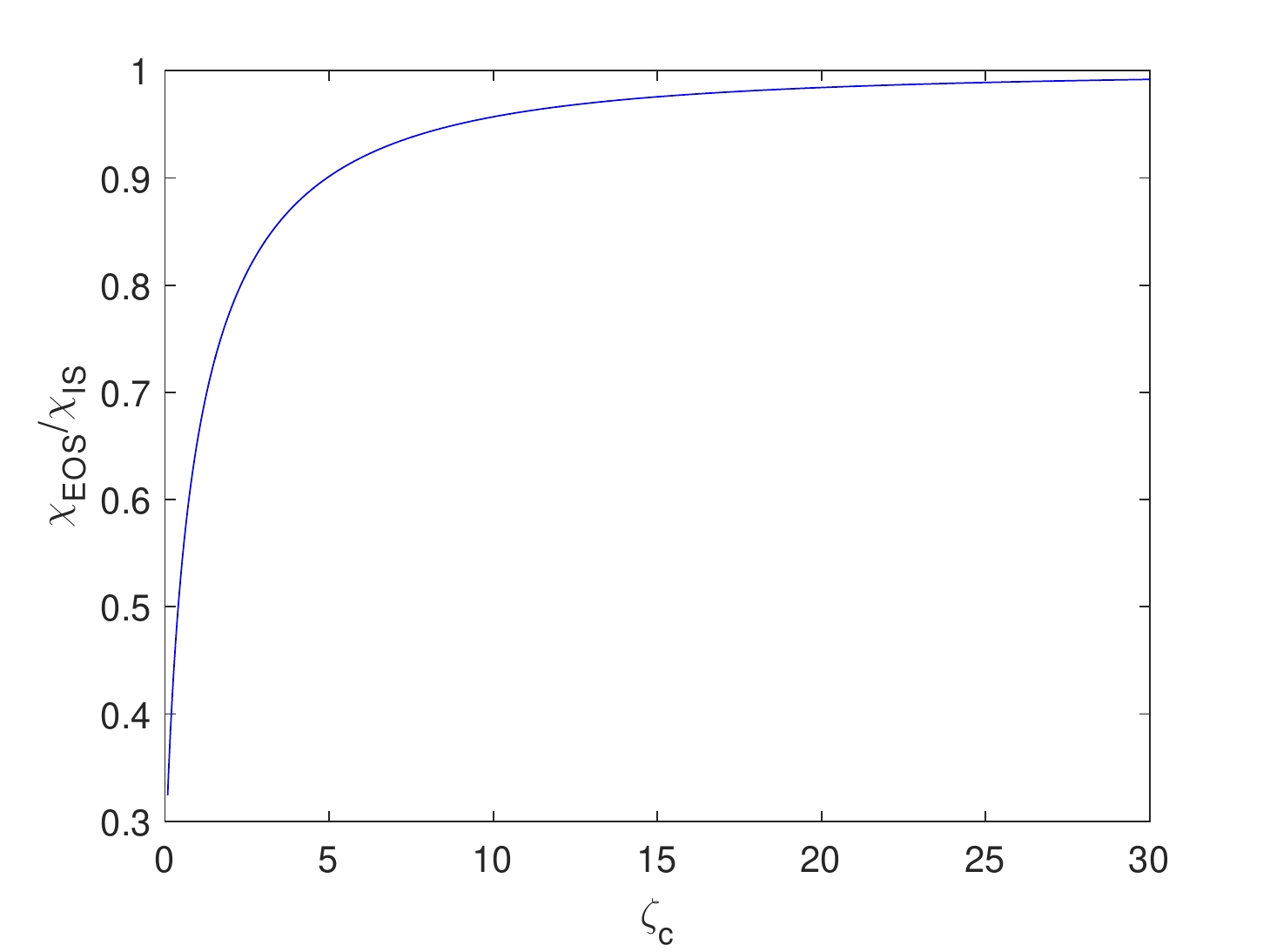} }
	\caption{Upper panel: plot of the factor $mn\chi$ as a function of $\zeta_c$ according to the quasi-equilibrium equation of state \eqref{equazdistato} (blue line) and to Israel and Stewart (red line). Lower panel: plot of the ratio between the two. We see that when the gas becomes relativistic ($\Theta \gtrsim m$) the convexity coefficient $\chi$ calculated in \citet{Israel_Stewart_1979} diverges from the results obtained with the quasi-equilibrium equation of state.}
	\label{fig:comp}
\end{figure}

In the first panel of figure \ref{fig:comp} we can see the comparison between the two predictions for $mn\chi$. In the second panel we show their ratio. We compare $mn\chi$ because they are dimensionless and depend only on $\zeta_c$. As can be seen from the figure, Israel-Stewart's prediction is always larger than the one of the quasi-equilibrium equation of state, but they become equal for low temperatures.

\subsection{Comparison}\label{4TEOSFARDMGOOE}

In this subsection we compare our kinetic approach with the Grad 14-moment approximation of \citet{Israel_Stewart_1979}, with the aim of explaining the behaviour of the second panel of figure \ref{fig:comp}.

Following the steps of \cite{Israel_Stewart_1979}, we focus on the function $\sigma$ we introduced in equation \eqref{sigma}, which for $\iota=0$ (non-degenerate limit) becomes
\begin{equation}
\sigma= -\ln \bigg( \dfrac{f h_p^3}{g} \bigg)
\end{equation}
and, using \eqref{lanostraflaNOOOOSTRA}, is equal to
\begin{equation}
\sigma = -\alpha + \zeta_c \sqrt{1+ \dfrac{\lambda^2 \vect{p}^2}{m^2}}.
\end{equation}
Since our aim is to study $f$ near equilibrium, we impose
\begin{equation}
\lambda = 1+\delta \lambda
\end{equation}
and we expand in $\delta \lambda$. If we define
\begin{equation}
\sigma_0 :=  -\alpha + \zeta_c \sqrt{1+ \dfrac{\vect{p}^2}{m^2}},
\end{equation}
we find that
\begin{equation}\label{miomiomio}
\sigma = \sigma_0 + \dfrac{\zeta_c \vect{p}^2}{m \epsilon} \delta \lambda.
\end{equation}
On the other hand, \cite{Israel_Stewart_1979} postulate a dependence of $\sigma$ on $\vect{p}$ (for locally isotropic matter elements) of the form
\begin{equation}\label{ISISIS}
\sigma_{IS} = \sigma_0 + \vect{p}^2 \nu,
\end{equation}
where $\nu$ is a parameter describing the displacement from equilibrium. For $\zeta_c \gg 1$ only the low energy states are significantly explored, $\epsilon \approx m$, therefore equation \eqref{miomiomio} reduces to \eqref{ISISIS} with
\begin{equation}
\nu= \dfrac{\zeta_c \delta \lambda}{m^2}.
\end{equation}
Therefore we have shown that in the low temperature limit the two approaches coincide and this is reflected in the fact that the respective predictions for $\chi$ tend to be the same for $\zeta_c \longrightarrow + \infty$. On the other hand, when the fluid becomes relativistic, the assumption about the shape of the perturbation becomes relevant. Considering that for large $|\vect{p}|$ we observe two different asymptotic behaviours:
\begin{equation}
\sigma \sim \dfrac{\zeta_c \lambda}{m} |\vect{p}|  \spc \spc  \sigma_{IS} \sim \nu \, \vect{p}^2,
\end{equation}
as the temperature increases we expect the accordance to fail, as it can be seen in figure \ref{fig:comp}. Substantially, our formulation models directly the evolution of the particle momentum distribution, which in the high temperature limit differs from the ansatz of \cite{Israel_Stewart_1979}.

\subsection{The role of degeneracy}

\begin{figure}
\subfigure{
	\includegraphics[width=0.5\textwidth]{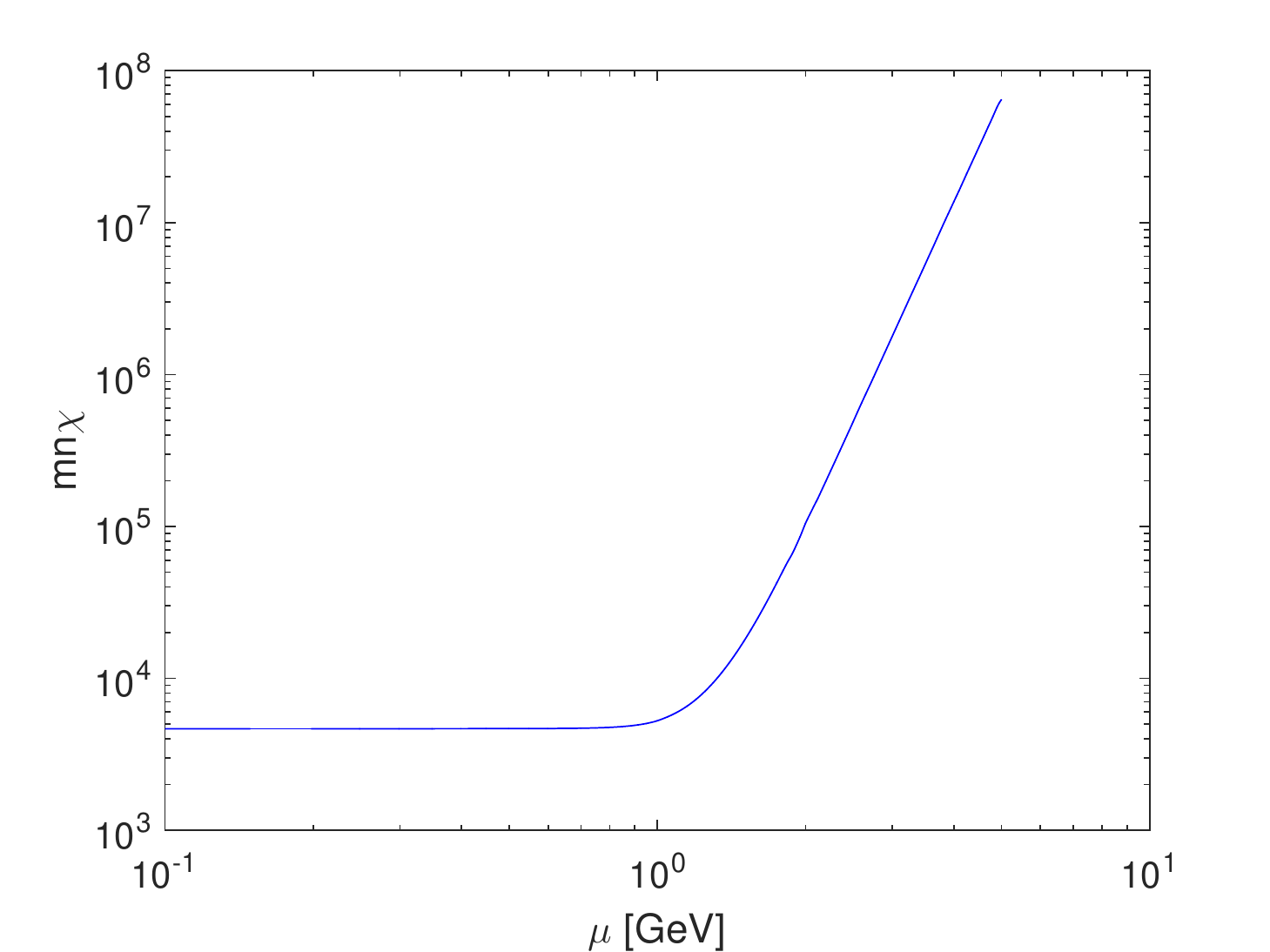}}
\subfigure{
	\includegraphics[width=0.5\textwidth]{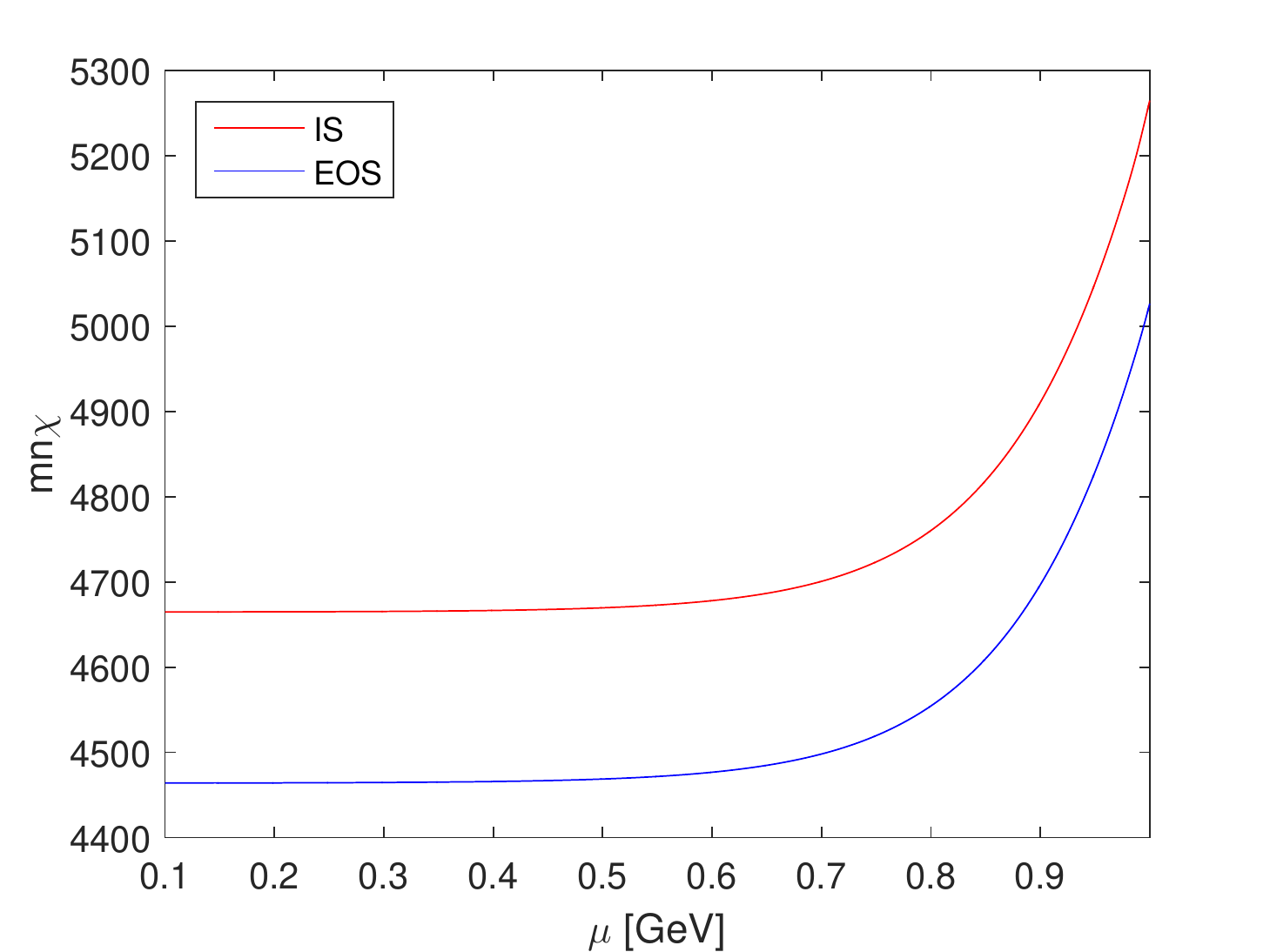}}
	\caption{Plot of $mn\chi$ as a function of the relativistic chemical potential. We are considering a gas of particles of mass 1 $GeV$, with a temperature $\Theta = 10^{12}$ K.}
	\label{fig:degenerato!}
\end{figure}

The calculations  in subsections \ref{TEOSFARDMGOOE}, \ref{2TEOSFARDMGOOE}, \ref{3TEOSFARDMGOOE} and \ref{4TEOSFARDMGOOE} can be performed also for a degenerate Fermi gas. 
However, since in the Fermi-Dirac case $\alpha$ cannot be factorised out in the integral expression for $n(\alpha,\zeta_c,\lambda)$, there seems to be no way of inverting this relation. This prevents us from approaching the problem analytically. Hence, we will present here only the plot of $mn\chi$ as a function of the chemical potential, see figure \ref{fig:degenerato!}. 

If $\mu < m$ we are in the non-degenerate limit and we recover the results of the previous subsections. In particular, in this limit $mn\chi$ does not depend on $\mu$, so we obtain the plateau that can be seen in figure \ref{fig:degenerato!} for low $\mu$. When $\mu>m$, however, $mn\chi \longrightarrow + \infty$. To understand this, consider the definition of the susceptibility given in equation \eqref{ssssusssccceptToooooo!}. In the variations performed to compute the second derivative one has to change only the shape of the distribution, keeping the same density of particles and entropy. However, in a degenerate gas, $f h_p^3/g$ is everywhere $1$ or 0, apart from a thin shell of momenta around the Fermi surface, whose thickness is proportional to the temperature. So only a small fraction of electrons near the Fermi momentum is involved in the variation. Since we are dividing by the whole energy per particles, $\mathcal{C} \longrightarrow 0$.  As a result, $mn\chi \longrightarrow +\infty$, see equation \eqref{berghione}.

In the degenerate limit the ratio $\chi_{EOS}/\chi_{IS} \longrightarrow 1$, which can be explained with a simple argument. Following \cite{Israel_Stewart_1979}, we introduce
\begin{equation}
\mathfrak{y}= \ln \bigg( \dfrac{f h_p^3/g}{1-f h_p^3/g} \bigg).
\end{equation}  
It is possible to check that in our theory
\begin{equation}
\mathfrak{y}= \alpha - \zeta_c \sqrt{1+ \dfrac{\lambda^2 p^2}{m^2}}.
\end{equation}
Analogously to what we did in the previous subsection we note that for small displacements from equilibrium we have
\begin{equation}
\mathfrak{y}= 
\alpha - \zeta_c \sqrt{1+ \dfrac{p^2}{m^2}} -\dfrac{\zeta_c p^2}{m \epsilon} \delta \lambda. 
\end{equation}
However in a degenerate gas the momenta which are involved moving out of equilibrium are a thin shell near the Fermi surface, so we can impose $p=p_F+q$, with $q$ small. So we can expand the last term and find
\begin{equation}
\mathfrak{y}= \alpha-\dfrac{\zeta_c p_F^2}{m \epsilon_F} \delta \lambda - \zeta_c \sqrt{1+ \dfrac{p^2}{m^2}}-\dfrac{\zeta_c \delta \lambda}{m} \dfrac{d}{dp}\bigg( \dfrac{p^2}{\epsilon} \bigg)\bigg|_{p_F} q .
\end{equation} 
Defining
\begin{equation}
\alpha_P := \alpha-\dfrac{\zeta_c p_F^2}{m \epsilon_F} \delta \lambda \spc \nu_P := -\dfrac{\zeta_c \delta \lambda}{m} \dfrac{d}{dp}\bigg( \dfrac{p^2}{\epsilon} \bigg)\bigg|_{p_F},
\end{equation}
we arrive at the form
\begin{equation}\label{concordinaoproprio}
\mathfrak{y}= \alpha_P - \zeta_c \sqrt{1+ \dfrac{p^2}{m^2}} + \nu_P q .
\end{equation}
On the other hand, \cite{Israel_Stewart_1979} assume a near-equilibrium distribution function of the form
\begin{equation}
\mathfrak{y}_{IS}= \alpha_{IS} -\zeta_c \sqrt{1+ \dfrac{p^2}{m^2}} + \nu p^2.
\end{equation}
Making the same expansion we obtain
\begin{equation}
\mathfrak{y}_{IS}= \alpha_{IS}+\nu p_F^2 -\zeta_c \sqrt{1+ \dfrac{p^2}{m^2}} + 2\nu p_F q.
\end{equation}
Then, noting that we can define
\begin{equation}
\alpha_P := \alpha_{IS}+\nu p_F^2  \spc \nu_P :=2\nu p_F,
\end{equation}
we recover \eqref{concordinaoproprio}. Considering that far from the Fermi surface $f h_p^3/g$ can be approximated to 0 or 1, we have shown that the prescription for the shape of the distribution function near equilibrium, in the degenerate limit, presented in \cite{Israel_Stewart_1979} coincides with ours. The only difference between the two approaches is given by the fact that we are using different charts, but the manifold $\mathcal{Z}$ is the same. Since all the physical quantities are independent from the initial chart which is employed in kinetic theory, the two theories are the same and $\chi_{EOS}/\chi_{IS} \longrightarrow 1$.

\begin{figure}
\begin{center}
	\includegraphics[width=0.5\textwidth]{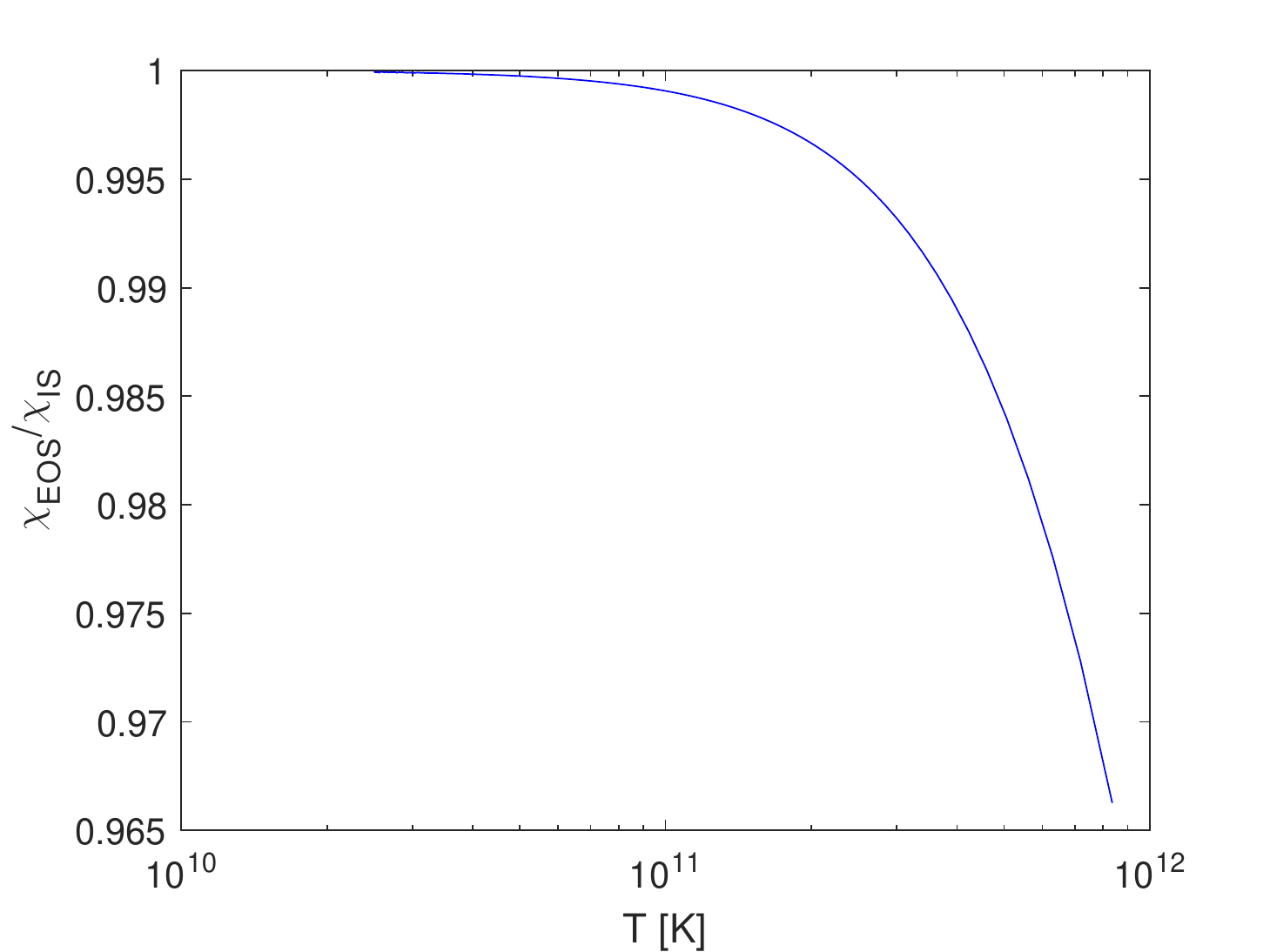}
	\caption{Plot of the ratio $\chi_{EOS}/\chi_{IS}$ as a function of temperature for a Fermion gas of particles with mass $m=1$ GeV. We consider the case $\mu \ll m$ (the plateau in figure \ref{fig:degenerato!}), because as the degeneration increases the function goes to 1.}
	\label{fig:plotdibryn}
	\end{center}
\end{figure}

In figure \ref{fig:plotdibryn} we show a plot of the behaviour of the ratio between the predictions of $\chi$ according to our equation of state and of \cite{Israel_Stewart_1979} for particles having the mass of the nucleons. To emphasize possible differences, the plot refers to the non-degenerate limit.

\section{From transport equations to bulk viscosity}

In the previous section we have shown how a kinetic description of an ideal gas can be used to compute the quasi-equilibrium equation of state. It is, however, possible to start directly from a relativistic transport equation and prove that the hydrodynamic description presented in section \ref{DHOLIF} can be derived from it. We report here the steps of this proof as a final demonstration of the coherence and universality of the approach. In addition, it will be evident that, at the level of kinetic theory, our approach is the direct result of a change of variables (the straightification of $W_{fr}$) in the formalism of Israel-Stewart, as can be verified through a direct comparison with \cite{Israel_Stewart_1979}. 

In order to simplify the equations we will work in the context of special relativity in Minkowskian coordinates.

\subsection{The continuity equations of kinetic theory} 

The evolution equation of the single-particle distribution (assuming absence of external forces) in a flat spacetime is \citep{cercignani_book}
\begin{equation}\label{vabenissimo!!!}
p^\nu \partial_\nu f = \mathfrak{C}[f],
\end{equation}
where $\mathfrak{C}$ is a collision term. Let us introduce the transport fluxes
\begin{equation}
\varphi_{(N)}^{\nu \alpha_1 ... \alpha_N} := \int p^{\alpha_1}...p^{\alpha_N} p^\nu f \dfrac{d_3 p}{p^0}
\end{equation}
and the collision tensors
\begin{equation}
\mathfrak{C}_{(N)}^{\alpha_1 ... \alpha_N}:= \int  p^{\alpha_1}...p^{\alpha_N} \mathfrak{C} \dfrac{d_3 p}{p^0}.
\end{equation}
They are totally symmetric tensors and, using \eqref{vabenissimo!!!}, satisfy
\begin{equation}\label{consSs}
\partial_\nu \varphi_{(N)}^{\nu \alpha_1 ... \alpha_N} = \mathfrak{C}_{(N)}^{\alpha_1 ... \alpha_N}.
\end{equation}
Note that $\varphi_{(0)}^\nu =n^\nu$ and $\varphi_{(1)}^{\nu \rho} =T^{\nu \rho}$ \citep{cercignani_book}. Conservation of particles and energy-momentum in the collisions imply 
\begin{equation}
\mathfrak{C}_{(0)}=0  \spc \mathfrak{C}_{(1)}^{\alpha_1}=0
\end{equation}
and \eqref{consSs} for $N=0$ and $N=1$ respectively become 
\begin{equation}
\partial_\nu n^\nu =0  \spc  \partial_\nu T^{\nu \rho}=0,
\end{equation}
in accordance with \eqref{continuity} and \eqref{energyconservation}. We can, now, introduce the tensors
\begin{equation}\label{DdD}
\begin{split}
& n_{(N)}^\nu := (-1)^N \varphi_{(N)}^{\nu \alpha_1 ... \alpha_N}  u_{\alpha_1}... u_{\alpha_N} \\
& T^{\nu \rho}_{(N)} := (-1)^{N-1} \varphi_{(N)}^{\nu \rho \alpha_2 ... \alpha_N}  u_{\alpha_2}... u_{\alpha_N}. \\
\end{split}
\end{equation}
It can be seen that
\begin{equation}
n_{(0)}^\nu = n^\nu  \spc  T_{(1)}^{\nu \rho} = T^{\nu \rho}.
\end{equation}
Since $\varphi_{(0)}^\nu$ has only one index, $T_{(0)}^{\nu \rho}$ is not well defined, but we extend the definition imposing 
\begin{equation}
T_{(0)}^{\nu \rho}:=0.
\end{equation}
Applying the divergence to the first equation of \eqref{DdD} we obtain the continuity equations
\begin{equation}\label{Tonino!!!!!!!}
\partial_\nu n_{(N)}^\nu = n c_N - N T^{\nu \rho}_{(N)} \partial_\nu u_\rho,
\end{equation}
where we have introduced the scalars
\begin{equation}
c_N= \dfrac{(-1)^N}{n} \mathfrak{C}_{(N)}^{\alpha_1 ... \alpha_N} u_{\alpha_1}...u_{\alpha_2}.
\end{equation}

\subsection{The local isotropy assumption}

Now we need to impose the condition that bulk viscosity is the only dissipative process occurring in the system. This, in section \ref{DHOLIF}, was encoded in the requirement of local isotropy in the reference frame of the fluid element. Therefore, in accordance with that approach we assume that in each point of the spacetime, $f$ is invariant under rotations in the reference frame defined by $n^\nu$. Then, it is immediate to show that, defined (from now on we work with $N>0$)
\begin{equation}\label{baluba}
\begin{split}
& \tilde{\mathcal{U}}_N := \dfrac{(-1)^{N+1}}{n} \varphi^{\nu \alpha_1 ... \alpha_N} u_\nu u_{\alpha_1}...u_{\alpha_N} \\
& \Psi_N := [T_{(N)}^{\nu \rho} g_{\nu \rho}+n \tilde{\mathcal{U}}_N]/3 , \\
\end{split}
\end{equation}
the tensors introduced in \eqref{DdD} have the form
\begin{equation}
\begin{split}
& n_{(N)}^\nu = \tilde{\mathcal{U}}_N n^\nu \\ 
& T^{\nu \rho}_{(N)} = [n\tilde{\mathcal{U}}_N + \Psi_N]u^\nu u^\rho + \Psi_N g^{\nu \rho}. \\
\end{split}
\end{equation}
This plugged into the continuity equations, we find
\begin{equation}\label{terRrmino!}
\dot{\tilde{\mathcal{U}}}_{N} = c_N -N \Psi_N \dot{v}.
\end{equation}
For $N=1$ we obtain \eqref{comparison}, since $\tilde{\mathcal{U}}_{(1)} = \tilde{\mathcal{U}}$ and $\Psi_{(1)} = \Psi$. The foregoing equation describes the evolution of the internal macroscopic degree of freedom $\tilde{\mathcal{U}}_N$, which is the particle average value of $(p^0)^N$ computed in the reference frame of the fluid. There are two contributions in the right-hand side: a collisional term, which represents the dissipative processes which tend to lead the system to local thermal equilibrium, and the direct effect of the expansion of the volume element.

\subsection{The thermodynamics of the fluid}\label{ttotffrtr}

It can be shown, see appendix \ref{TMOSPD}, that the infinite set of the $\tilde{\mathcal{U}}_N$, together with $n$, contains the same amount of information as $f$. On the other hand, in order for a hydrodynamical description of the fluid to be possible, we need to assume that only a finite number of them is necessary to know the macroscopic local state of matter. Therefore we can impose that 
\begin{equation}\label{ilprimodeiptizi}
(v,\tilde{\mathcal{U}}_N)_{N=1,...,l}
\end{equation}
is a global chart of the manifold of the thermodynamic states. This is equivalent to saying that all the $\tilde{\mathcal{U}}_N$ for $N>l$ can be written as functions of $v$ and the $\tilde{\mathcal{U}}_N$ with $N \leq l$. As $l$ grows we explore an hierarchy of more and more refined fluid theories. For $l=0$ we have the barotropic perfect fluid, for $l=1$ the finite temperature perfect fluid, for $l>1$ we have more and more complicated models of bulk viscosity and, in the limit $l \longrightarrow +\infty$ we recover full kinetic theory. 

We have written the manifold of states in the form \eqref{charttt}, where the $\tilde{\mathcal{U}}_N$ for $N>1$ play the role of the $\alpha_A$. Now we need to introduce the vector field $W_{fr}$. In the chart \eqref{ilprimodeiptizi} it can be decomposed as
\begin{equation}\label{expanDo}
W_{fr} = \dfrac{\partial}{\partial v} + \sum_{N=1}^l W_{fr}(\tilde{\mathcal{U}}_N) \dfrac{\partial}{\partial \tilde{\mathcal{U}}_N}.
\end{equation}
Its action on a thermodynamic variable gives the derivative of the variable with respect to the volume per particle along the curve drawn by the system in an expansion which is faster than the relaxation time-scale, see subsection \ref{GEOS}. On the other hand, we show in appendix \ref{sec:adexp} that the aforementioned curve can be described in a parametric way, in the context of kinetic theory of weakly interacting gases, through the condition \eqref{transformation}.\footnote{
    We are working in the reference frame of the fluid element here.
} Therefore, using the chain rule through equation \eqref{xcv}, we can write
\begin{equation}
W_{fr}(\tilde{\mathcal{U}}_N) = \dfrac{1}{3v} \dfrac{d\tilde{\mathcal{U}}_N(\lambda)}{d\lambda} \bigg|_{\lambda=1}.
\end{equation} 
Using the first definition in \eqref{baluba} and the fact that we are in the reference frame of the matter element, we find
\begin{equation}
W_{fr}(\tilde{\mathcal{U}}_N) = -\dfrac{N}{3}  \int (m^2 + \vect{p}^2)^{N/2-1} \vect{p}^2 f(\vect{p}) d_3 p.
\end{equation} 
On the other hand it is possible to verify that
\begin{equation}
\Psi_N = \dfrac{1}{3}  \int (m^2 + \vect{p}^2)^{N/2-1} \vect{p}^2 f(\vect{p}) d_3 p,
\end{equation}
hence we obtain
\begin{equation}\label{hHHHHHHHh}
W_{fr}(\tilde{\mathcal{U}}_N) = -N \Psi_N.
\end{equation}
Note that, remembering equation \eqref{KkK}, the above formula reduces to \eqref{grandpotepress} for $N=1$.

Plugging \eqref{hHHHHHHHh} into \eqref{terRrmino!} we obtain
\begin{equation}\label{mammabusnelin}
\dot{\tilde{\mathcal{U}}}_{N} = c_N +W_{fr}(\tilde{\mathcal{U}}_N) \dot{v}.
\end{equation}
If we impose the collision term to be zero, the above equation becomes
\begin{equation}
\dfrac{d \tilde{\mathcal{U}}_N}{d v} \bigg|_{along \,a \, worldline} = W_{fr}(\tilde{\mathcal{U}}_N),
\end{equation}
which means that in the absence of dissipation the thermodynamic variables change along the worldlines according to the transformation \eqref{transformation}, produced by the expansions and contractions of the volume element.

\subsection{The chemical-like chart}

Since the chart \eqref{ilprimodeiptizi} has the form \eqref{charttt} and we have the vector field $W_{fr}$, we can introduce a chart of the type $(v,x_s,x_A)$,\footnote{Recall that $W_{fr}(x_s)=0$ is guaranteed to be true because of equation \eqref{qwertyuioopljhggf}, see appendix \ref{sec:adexp} for more details. Furthermore, equation $\dot{x_s} \geq 0$ is a necessary product of Boltzmann's H-theorem \citep{cercignani_book}} see subsection \ref{GEOS}, with $x_A=x_A(v,\tilde{\mathcal{U}}_N)$ dimensionless variables satisfying the condition
\begin{equation}
W_{fr}(x_A) = 0.
\end{equation}
Using the expansion \eqref{expanDo}, we find
\begin{equation}\label{annullamentototale}
\dfrac{\partial x_A}{\partial v} + \sum_{N=1}^l W_{fr}(\tilde{\mathcal{U}}_N) \dfrac{\partial x_A }{\partial \tilde{\mathcal{U}}_N}=0,
\end{equation}
where the partial derivatives are referred to the chart \eqref{ilprimodeiptizi}.

We can compute the variation of $x_A$ along the worldlines of the matter elements. Using the chain rule we can write
\begin{equation}
\dot{x}_A = \dfrac{\partial x_A}{\partial v} \dot{v} + \sum_{N=1}^l \dfrac{\partial x_A}{\partial \tilde{\mathcal{U}}_N} \dot{\tilde{\mathcal{U}}}_N,
\end{equation}
with the aid of equations \eqref{mammabusnelin} and \eqref{annullamentototale} we obtain
\begin{equation}\label{CconA}
\dot{x}_A = c_A,
\end{equation}
with
\begin{equation}
c_A = \sum_{N=1}^l \dfrac{\partial x_A}{\partial \tilde{\mathcal{U}}_N} c_N.
\end{equation}
In the absence of collisions (dissipation), $c_A$ vanishes, so $x_A$ is constant along the worldline, recovering the frozen limit. This proves, directly from kinetic theory, since it is a consequence of the continuity equations \eqref{Tonino!!!!!!!}, that it is always possible to find $l$ coordinates out of $l+1$ thermodynamic degrees of freedom which are altered only by the relaxation processes.  

The final step we need to make consists of showing that, near equilibrium, equation \eqref{CconA} can be rewritten in the form \eqref{strongG}. However this is not a hard task, in fact we can write the $c_A$ in the chart $(v,x_s,\mathbb{A}^A)$ and expand near equilibrium (for small $\mathbb{A}^A$):
\begin{equation}
c_A(v,x_s,\mathbb{A}^B) =  c_A(v,x_s,0) + \dfrac{\partial c_A}{\partial \mathbb{A}^B}(v,x_s,0) \mathbb{A}^B.
\end{equation}
Since the collision term $\mathfrak{C}$ vanishes in equilibrium we need to impose $c_A(v,x_s,0)=0$. Defining
\begin{equation}
\Xi_{AB}=n \dfrac{\partial c_A}{\partial \mathbb{A}^B}(v,x_s,0),
\end{equation} 
we conclude the proof.

\subsection{The explosion of the degrees of freedom and the universality of the Navier-Stokes equation}\label{explofdegree}


At the beginning of section \ref{ttotffrtr} we assumed that only a finite number of $\tilde{\mathcal{U}}_N$ were independent. This is a necessary condition to make the hydrodynamic description possible and is one of the fundamental assumptions invoked in sections \ref{TOOOEF}, \ref{DHOLIF} and \ref{eobv!!!}. 
Although this may be a good approximation already for $l=2$ (depending on the shape of the collision integral), this may not be always the case: we cannot exclude a priori that a large amount of variables are  required to describe the fluid. 
The condition $l \rightarrow + \infty$ implies that there are infinite telegraph-type equations \eqref{zazza}, making the hydrodynamic problem unsolvable. This pathological explosion of degrees of freedom is  a symptom of the issue pointed out in subsection \ref{ottopuntouno}: in this system the existence of a hydrodynamic description is guaranteed only for $\tau_H \gg \tau_M$.

This problem marks a fundamental difference between the hyperbolic and the parabolic case. 
As pointed out in section \ref{paraparaparapara}, in the hyperbolic case a large $l$ implies more independent variables and, therefore, an increasing complexity. 
On the other hand, in the parabolic limit all the contributions add up, see equation \eqref{zetta}, in the coefficient $\zeta$ which, alone, contains all the information about the thermodynamics and kinetics of the substance. For this reason, even if $l \rightarrow +\infty$, in the parabolic limit the hydrodynamic theory is still possible, making Navier-Stokes a universal equation. 
Thus, a weakly interacting gas, whose bulk viscosity is produced by the relaxation of the momentum distribution to equilibrium, is a perfect example of a fluid which is efficiently described through Navier-Stokes, but which in the hyperbolic regime may need an $l$ larger than 2, in contrast with what has been done up to now, as discussed in subsection \ref{Qes}.

\section{Conclusions}

We derived, by application of the principles of extended irreversible thermodynamics, a general hydrodynamic description for bulk-viscous fluids in General Relativity, that builds on the covariant multifluid formalism of Carter and collaborators \citep{carter1991, carter94, carter1995} and naturally allows for a symmetric hyperbolic form of the system \citep{Causality_bulk} and causal solutions. 
The present theory is thus well suited for numerical applications.

Our model is not based on near-equilibrium assumptions, but only a time-scale separation between the equilibration of a restricted (tractable) number of macroscopic degrees of freedom and all the microscopic ones. We have proven that this can always be recast into a multi-component single fluid with a number of currents which grows with the number of out-of-equilibrium degrees of freedom.

In this context, dissipation is modelled as chemical transfusion, even in the absence of real chemical reactions. 
The chemically-induced bulk viscosity in neutron stars has been shown to be the simplest particular case of our description, and the dependence of the effective bulk viscosity coefficient on the frequency, generally derived from a perturbative approach \citep{Sawyer_Bulk1989}, has been shown to arise directly from the telegraph-type form equations in our model. Therefore our approach provides the machinery to take any equation of state and set of reaction rates (e.g. in the neutron star interior) and construct a hyperbolic model for bulk viscosity. In such a context the standard bulk viscosity coefficient itself can also be recovered by taking the parabolic (low frequency) limit of the theory.


The model (for bulk viscosity) of \citet{Israel_Stewart_1979} has been proven to emerge as an expansion near equilibrium of our general approach, and we give a formula for its coefficients in the chemically-induced case. Varying the dynamical time-scale of the hydrodynamics, different regimes of the theory can appear, such as the fast limit in which the new degrees of freedom appear to be frozen on the hydrodynamic time-scale, and the slow limit in which the parabolic Navier-Stokes equations are recovered. 

The approach has been, finally, employed in the context of the kinetic theory of ideal gases of fermions. This has provided a  proof of the fact that bulk viscosity vanishes in the non-relativistic and in the ultra-relativistic limit. In the intermediate case we have derived the equation of state for the out-of-equilibrium gas in a minimal model with $l=2$ (i.e. allowing for the equation of state to depend on temperature and an additional variable), and calculated the bulk viscosity coefficients. In the degenerate and cold limit our results coincide with those derived from the Grad 14-moment approximation of \cite{Israel_Stewart_1979}. At higher temperatures, for $T\gtrsim 1$ MeV (i.e. $T\gtrsim 10^{10}$ K) however, while the behaviour is qualitatively similar, there are quantitative differences between the two approaches, as our formulation models directly the evolution of the particle momentum distribution, which in the high temperature limit differs from the ansatz of \cite{Israel_Stewart_1979}.

The results of this paper are applicable to a number of astrophysical problems in which a general relativistic description of viscous matter is needed, including numerical studies of a binary neutron star merger and its remnant, but also for the damping of modes of oscillation of compact as well as classical stars. Indeed, we would like to remark that they have recently been used in the context of radiation hydrodynamics, to model the dissipation which is generated from the interaction between matter and radiation in the limit of infinite elastic-scattering opacity \citep{gavassino2020radiation}. 

Future directions of investigation concern the possibility of relaxing the assumption of isotropy of the fluid element. This naturally leads one to the inclusion of heat conduction, charge conductivity and shear viscosity in the theory. In addition, relaxing the local rotation-invariance assumption is a necessary step for a consistent implementation of superfluidity. Unfortunately, the straightification technique used in subsection \eqref{GEOS} to simplify the formulation of the dynamics cannot be applied to anisotropic fluid elements, as expansions in orthogonal directions may not commute
with each other (preventing a simultaneous straightification). As a consequence, within the Extended Irreversible Thermodynamics framework discussed here, there is a large variety of different models for shear viscosity that have the same number of degrees of freedom \citep{Hishcock1983,Romatschke2010,Shibata2017}, contrarily to what happens for pure bulk viscosity.

\section*{Acknowledgements}

We acknowledge support from the Polish National Science Centre grants SONATA BIS 2015/18/E/ST9/00577 and OPUS 2019/33/B/ST9/00942. Partial support comes from PHAROS, COST Action CA16214. The authors thank D. Hilditch and N. Andersson for reading
the manuscript and providing critical comments. 

\appendix

\section{Mathematical calculations}

In this appendix we present the proofs which were removed from the main body of the paper.

\subsection{Straightifying the generator of the fast expansions}\label{STGOTFE}

Let us choose an arbitrary reference volume $v_0$ and define $\mathcal{Z}_{v_0}$ to be the $l$-dimensional submanifold of $\mathcal{Z}$ given by the condition $v=v_0$. It represents the set of all the possible quasi-equilibrium states in which the fluid can be found when the opposite walls of the box have a distance $L=(Nv_0)^{1/3}$. Let $(x_s,x_A)$, $A=1,...,l-1$, be a chart of $\mathcal{Z}_{v_0}$. We have chosen one coordinate to be the entropy per particle, while the remaining ones can be chosen freely. For later symbolic and interpretative convenience we impose the variables $x_A$ to be dimensionless. 
Now, given an arbitrary point $p$ in $\mathcal{Z}$ with a given value of the coordinate $v$, we can imagine to perform an expansion (or contraction) of the volume in a time-scale $\tau_{fr}$ such that at the end of the process $v=v_0$, so the system will occupy a point of $\mathcal{Z}_{v_0}$, with coordinates $(x_s,x_A)$. Thus we can define the map from $\mathcal{Z}$ to $\mathbb{R}^{l+1}$
\begin{equation}\label{laCartaCartona}
p  \quad \longmapsto \quad  (v,x_s,x_A).
\end{equation}
Note that $x_s$ coincides with the entropy per particle of $p$, because it is conserved along the transformation. Since in the transformation no entropy production occurs, the process is reversible, which means that the opposite transformation $v_0 \rightarrow v$ cannot be distinguished from the time-reversed of the transformation $v \rightarrow v_0$. This implies that if we make the cycle of expansion and contraction $v \rightarrow v_0\rightarrow v$ in a time-scale $\tau_{fr}$ we should end in the initial state. Thus two distinct points $p$ and $p'$ with the same $v$ have to be sent, in the transformation $v \rightarrow v_0$, into two distinct points of $\mathcal{Z}_{v_0}$, implying that the map \eqref{laCartaCartona} is one-to-one and represents a global chart of $\mathcal{Z}$.

We have defined a convenient coordinate system in which the curves explored making fast expansions are given by the conditions $x_s =const$  and $x_{A}=const$. To see this it is sufficient to note that two points $p$ and $p'$ belonging to the same curve, with different volumes $v$ and $v'$, end up in the same point of $\mathcal{Z}_{v_0}$ when their respective volumes are sent into $v_0$. Thus they have the same value of $x_s$ and $x_A$. 

The generator of the curves $x_s,x_{A}=const$ is the vector field
\begin{equation}
W_{fr} := \dfrac{\partial}{\partial v} \bigg|_{x_s,x_A} \, ,
\end{equation}
which, applied to any thermodynamic function, describes its variation in an expansion occurring in a time-scale $\tau_{fr}$ and therefore must coincide with $W_{fr}$ introduced in \eqref{GEOS}.

\subsection{The invertibility of the relaxation-time matrix}\label{TIOTTM}

We will prove the invertibility of $\delta\indices{^A _B} + i \omega \tau\indices{^A _B}$ with a reduction ad absurdum. Let us assume that there is a non-zero complex vector $p^B$ such that
\begin{equation}\label{jutro}
(\delta\indices{^A _B} + i \omega \tau\indices{^A _B})p^B =0.
\end{equation}
Then we have that
\begin{equation}\label{juftre}
\tau\indices{^A _B} \tau\indices{^B _C} p^C =- \omega^{-2} p^A.
\end{equation}
Now consider the differential \eqref{dG}. Since it describes an exact 1-form, the equivalence between the second mixed derivatives produces the Maxwell relation
\begin{equation}
\dfrac{\partial x_A}{\partial \mathbb{A}^B} \bigg|_{v,x_s} = \dfrac{\partial x_B}{\partial \mathbb{A}^A} \bigg|_{v,x_s} .
\end{equation}
Using \eqref{inversemetric} and \eqref{TTau}, this implies
\begin{equation}
\Xi_{AC} \, \tau\indices{^C _B} = \tau\indices{^C _A} \Xi_{CB}.
\end{equation}
Contracting with $(p^A)^* \tau\indices{^B _D} p^D$ we obtain
\begin{equation}
 \Xi_{AC} (p^A)^* (\tau\indices{^C _B} \tau\indices{^B _D} p^D) = (p^A)^* \tau\indices{^C _A} \Xi_{CB} \, \tau\indices{^B _D} p^D,
\end{equation}
which, using \eqref{jutro} and \eqref{juftre}, becomes
\begin{equation}
- \Xi_{AC} (p^A)^* \, p^C = \Xi_{CB} (  p^C)^* \, p^B.
\end{equation}
Considering that $\Xi_{AB}$ is a symmetric positive definite matrix and $p^A$ is different from zero we have a contradiction.

\subsection{Bulk viscosity of non(ultra)-relativistic ideal gases}\label{staiqui}

We have constructed the coordinates $x_A$ in a way that $W_{fr}(x_A)=0$, see equation \eqref{Wad}. Expressing this condition in the chart $(v,x_s,\mathbb{A}^B)$ we find
\begin{equation}\label{poiuyt}
\dfrac{\partial x_A}{\partial v} \bigg|_{\mathbb{A}^B, x_s} + \sum_{B=1}^{l-1} \dfrac{\partial x_A}{\partial \mathbb{A}^B} \bigg|_{v,x_s} W_{fr}(\mathbb{A}^B) =0. 
\end{equation}
The requirement that the curves generated by $W_{fr}$ starting in the equilibrium surface (given by the conditions $\mathbb{A}^B =0$) are entirely contained in the equilibrium surface itself is expressed by the 
tangenciality condition $W_{fr}(\mathbb{A}^B) =0$ $\forall B$ whenever all the $\mathbb{A}^B$ vanish, leading to
\begin{equation}\label{iovincoancora}
\dfrac{\partial x_A}{\partial v} \bigg|_{\mathbb{A}^B=0, x_s} =0.
\end{equation}
As it can be seen from equation \eqref{qwerty}, the condition \eqref{iovincoancora} immediately implies that $\Pi$ vanishes. Furthermore, one can easily verify from equation \eqref{ilgonzo} that if a fluid element is prepared in a local equilibrium state, then $\mathbb{A}^B=0$ $\forall B$ along its worldline, provided that \eqref{iovincoancora} holds. Thus we have proven that the bulk viscosity is zero whenever $W_{fr}(\mathbb{A^A})=0$ in equilibrium. However this must always be true for non-relativistic and ultra-relativistic gases. In fact, starting from \eqref{EOS}, one can immediately prove the Maxwell relation
\begin{equation}\label{dfghjk}
\dfrac{\partial \mathbb{A}^A}{\partial v} \bigg|_{x_s,x_B} = \dfrac{\partial \Psi}{\partial x_A} \bigg|_{v,x_s} .
\end{equation}
However for a non(ultra)-relativistic gas we have that, independently from the fact that the fluid element is in thermodynamic equilibrium or not,
\begin{equation}\label{politr}
\Psi = (\Gamma-1)(\mathcal{U}-mn).
\end{equation}
This is a kinetic identity which holds for any isotropic $f$. Note that in the ultra-relativistic case one has to formally set $m=0$. 
Using \eqref{politr}, equation \eqref{dfghjk} becomes
\begin{equation}
W_{fr}(\mathbb{A}^A) = -(\Gamma -1)n\mathbb{A}^A  ,
\end{equation}
which implies that in equilibrium
\begin{equation}
W_{fr}(\mathbb{A}^A) =0.
\end{equation}

\subsection{Affinity of a diluted relativistic gas}\label{Aoadrg}

According to \eqref{EOS} the affinity associated to $\zeta_c$ is
\begin{equation}
\mathbb{A} = - \dfrac{\partial \tilde{\mathcal{U}}}{\partial \zeta_c} \bigg|_{v,x_s} = -\dfrac{\partial \tilde{\mathcal{U}}}{\partial \zeta_c} \bigg|_{\lambda} - \dfrac{\partial \tilde{\mathcal{U}}}{\partial \lambda} \bigg|_{\zeta_c} \dfrac{\partial \lambda}{\partial \zeta_c} \bigg|_{v,x_s} .
\end{equation}
Using the results of section \ref{abcdefghi}, it is possible to check that the affinity can be written as the sum of three terms:
\begin{equation}\label{zmalsjdhfhtyieo}
\mathbb{A} = \mathbb{A}_{(1)} + \mathbb{A}_{(2)} + \mathbb{A}_{(3)},
\end{equation} 
with
\begin{equation}
\begin{split}
& \mathbb{A}_{(1)} = \tilde{\mathcal{U}} \phi' \\
& \mathbb{A}_{(2)} = m e^{-\phi} \int_0^{+\infty} \xi^2 e^{-\zeta_c \sqrt{1+\xi^2}} \sqrt{1+\xi^2} \sqrt{1+ \dfrac{\xi^2}{\lambda^2}} d\xi \\
& \mathbb{A}_{(3)} =  3\Theta \dfrac{\partial}{\partial \zeta_c} (\ln \lambda)\bigg|_{v,x_s} = - \Theta \zeta_c \phi'' . \\
\end{split} 
\end{equation}
Now we only need to check that for $\lambda =1$ we find $\mathbb{A}=0$. Note that for $\lambda=1$
\begin{equation}\label{qmelyl}
  \tilde{\mathcal{U}} = - m \phi'  \spc \Theta =  \dfrac{m}{\zeta_c} 
\end{equation}
and
\begin{equation}
\spc \mathbb{A}_{(2)} =  m (\phi'' + {\phi'}^2).
\end{equation}
When we plug these formulas into the expression for $\mathbb{A}$ we find that the three terms cancel out, leaving
\begin{equation}
\mathbb{A} = 0,
\end{equation}
which is what we wanted to prove.

\subsection{The second derivative of the energy in the generalised reaction coordinate}\label{TSDOTEITGRC}

Using \eqref{zmalsjdhfhtyieo} and the definition of $\mathbb{A}$ we see that we need to compute
\begin{equation}\label{derivoibruttoenontrovoilbello}
\dfrac{\partial^2 \tilde{\mathcal{U}}}{\partial \zeta_c^2}  \bigg|_{v,x_s} =- \dfrac{\partial \mathbb{A}_{(1)}}{\partial \zeta_c} \bigg|_{v,x_s} - \dfrac{\partial \mathbb{A}_{(2)}}{\partial \zeta_c} \bigg|_{v,x_s} - \dfrac{\partial \mathbb{A}_{(3)}}{\partial \zeta_c} \bigg|_{v,x_s}.
\end{equation}
However we need to compute the result only in equilibrium, so we can use this fact to simplify the formulas, for example
\begin{equation}
\dfrac{\partial \mathbb{A}_{(1)}}{\partial \zeta_c} \bigg|_{v,x_s} = -\mathbb{A} \phi' + \tilde{\mathcal{U}} \phi'' = -m \phi' \phi''. 
\end{equation}
We have employed the fact that in equilibrium the affinity vanishes and that the first equation of \eqref{qmelyl} holds. 

With direct calculations one can show that in equilibrium
\begin{equation}
\dfrac{\partial \mathbb{A}_{(2)}}{\partial \zeta_c} \bigg|_{v,x_s} = m \phi''' +2m \phi' \phi'' + \dfrac{m \zeta_c \phi''}{3} (\phi'' + {\phi'}^2 - 1)
\end{equation}
and
\begin{equation}
\begin{split}
& \dfrac{\partial \mathbb{A}_{(3)}}{\partial \zeta_c} \bigg|_{v,x_s} = -m \phi'''+  m \phi' \phi''  \\
&  - \dfrac{m\phi''}{\zeta_c} +   \dfrac{m \zeta_c \phi''}{3} ( {\phi'}^2 - 1) -  \dfrac{mJ}{3} (\zeta_c \phi'')^2 . \\
\end{split}
\end{equation}
Plugging these results into \eqref{derivoibruttoenontrovoilbello} we find
\begin{equation}
\dfrac{\partial^2 \tilde{\mathcal{U}}}{\partial \zeta_c^2}  \bigg|_{v,x_s} =
 m \phi'' \bigg[\dfrac{1}{\zeta_c}  -2 \phi'     -\dfrac{ \zeta_c }{3} (\phi'' + 2{\phi'}^2 - 2) +   \dfrac{J \zeta_c^2 \phi''}{3} \bigg] .
\end{equation}
Using the formulas of \eqref{conversion} it can be further simplified to reach the form
\begin{equation}
\dfrac{\partial^2 \tilde{\mathcal{U}}}{\partial \zeta_c^2}  \bigg|_{v,x_s} = \dfrac{m \zeta_c {\phi''}^2}{3} \bigg( 1+ \zeta_c J - \dfrac{3}{\zeta_c^2 \phi''} \bigg).
\end{equation} 

\subsection{The moments of single-particle distribution}\label{TMOSPD}

Working in the reference of the matter element we have that $f$ can be seen as a function of $p^0$. In particular we can introduce
\begin{equation}
I :=\dfrac{dn}{dp^0} =4\pi |\vect{p}| p^0 f ,
\end{equation}
the density of particle per unit energy. Clearly it contains the same amount of information of $f$ and it is easy to verify that
\begin{equation}\label{WWWWwwwwWWWWW}
 n = \int_0^{+\infty} I d p^0 \spc
 n\tilde{\mathcal{U}}_N = \int_0^{+\infty} (p^0)^N I d p^0. 
\end{equation}
Now let us study the series
\begin{equation}
F(z) := \sum_{N=0}^{+\infty} \dfrac{(-z)^N}{N!} n\tilde{\mathcal{U}}_N,
\end{equation}
where $z$ is an arbitrary complex number with positive real part. Under this condition we can bring the series in the integrals presented in \eqref{WWWWwwwwWWWWW}, giving
\begin{equation}
F(z) = \int_0^{+\infty} e^{-zp^0} I(p^0) d p^0.
\end{equation}
This is the Laplace transform of $I$, which can be inverted, so we arrive at the formula
\begin{equation}
I = \mathfrak{L}^{-1}\{ F \} (p^0),
\end{equation}
where $\mathfrak{L}^{-1}$ is the inverse Laplace transform. This proves that, if $n$ and all the $\tilde{\mathcal{U}}_N$ are known, then it is possible to reconstruct $I$ and therefore $f$.

\section{Absence of a macroscopic criterion to constrain the temperature}\label{statmec}

In this appendix we explain the physical origin of the ambiguity in the definition of the temperature exposed in subsection \ref{gauge}.

Let us consider the homogeneous fluid in the box described in subsections \ref{Qes} and \ref{GEOS}. If we take the limit
\begin{equation}
\tau_M \longrightarrow + \infty ,
\end{equation}
the variables $x_A$ can be considered constants of motion. We are essentially switching off the microscopic processes at the origin of their relaxation. Imagine in this limit to put the system in contact with an ideal constant temperature heat bath, with temperature $\Theta_H=const$. This is an effectively infinite mass-energy reservoir characterized by the equation of state
\begin{equation}
E_H (S_H) = E_{H0} + \Theta_H S_H,
\end{equation}
where $E_H$ is the energy measured in the frame of the walls. The second principle of thermodynamics, combined with the conservation of the total energy, tells us that after an equilibration process the substance in the box will reach the state which minimizes the quantity
\begin{equation}\label{dacapoancoraallinizio}
F = E - \Theta_H S.
\end{equation}
Therefore, considering that $N$ is assumed fixed (no exchange of particles happens), we find that we have to impose 
\begin{equation}
\delta \tilde{\mathcal{F}} = \dfrac{\delta F}{N} = (\Theta - \Theta_H)\delta x_s + \mathcal{K} \delta v  - \sum_{A=1}^{l-1} \mathbb{A}^A \delta x_A =0.
\end{equation} 
Imposing that the walls are fixed ($\delta v =0$) and that the interaction with the bath does not destroy the conservation of $x_A$ ($\delta x_A =0$), the only condition we get is
\begin{equation}
\Theta = \Theta_H .
\end{equation}
When a system is in thermal equilibrium with a bath, its temperature must coincide with $\Theta_H$, so we can interpret $\Theta$ as the generalization of the notion of temperature to quasi-equilibrium states. 

Now let study the evolution of the variables $y_B$, introduced in \eqref{qqq}, during the process of equilibration with the bath. Considering that $v$ and $x_A$ are constant, we find
\begin{equation}\label{Thermo}
\dfrac{dy_B}{dt} = \dfrac{\partial y_B}{\partial v} \dfrac{dv}{dt} + \dfrac{\partial y_B}{\partial x_s} \dfrac{d x_s}{dt} +\sum_{A=1}^{l-1} \dfrac{\partial y_B}{\partial x_A} \dfrac{dx_A}{dt} = \dfrac{\partial y_B}{\partial x_s} \dfrac{d x_s}{dt}.
\end{equation}
Now let us assume that the $y_B$ are built in a way that $\Theta \neq \Theta'$, see \eqref{gaugiandotutto}. Then there must be a $B$ such that
\begin{equation}
\dfrac{\partial y_{B}}{\partial x_s} \neq 0,
\end{equation}
so, in general, during the evolution we will have
\begin{equation}
\dfrac{dy_B}{dt} \neq 0.
\end{equation}
This reveals where the ambiguity in the definition of the temperature comes from: since the interaction with the bath modifies the values of energy and entropy of the fluid, if we assume that the $x_A$ are constant we will necessarily find that some of the $y_B$ are no more conserved and vice versa. Therefore the ambiguity in the definition of the temperature reflects the fact that there is no  macroscopic criterion to decide which set of possible variables $x_A$ is conserved in the interaction with an ideal heat bath. This kind of problem has implications also in the definition of the equilibrium temperature in a relativistic context as it is the fundamental origin of the famous Planck-Ott imbroglio \citep{gavassino2020zeroth}. 

To solve the ambiguity, if possible, one has to study the dissipative processes involved in the relaxation of the system to thermodynamic equilibrium and understand if there is a particular choice of the variables $x_A$ which is more reasonable to be conserved in an interaction with an ideal heat bath. This is a problem of kinetic theory which goes beyond a pure thermodynamic description. 

Let us assume that an unambiguous criterion to determine whether a state variable is conserved in the interaction with the bath exists and that there are two alternative complete sets of $l-1$ coordinates, $x_A$ and $y_B$ (conserved in the adiabatic expansions), satisfying this criterion. Then combining the constraint \eqref{aDDib} with the fact that the left-hand side of \eqref{Thermo} must be equal to zero, we find that
\begin{equation}\label{gaugeperfetto}
y_B = y_B (x_A),
\end{equation} 
which implies
\begin{equation}\label{invv}
\mathcal{K} = \mathcal{K}'  \spc  \Theta = \Theta',
\end{equation}
removing all the ambiguities. 

Note that, even in this case, the chemical potential is not gauge invariant out of equilibrium. In fact, from \eqref{euller}, it is easy to see that, if \eqref{invv} holds, then 
\begin{equation}
 \mu - \sum_{A=1}^{l-1} \mathbb{A}^A x_A = \mu' - \sum_{B=1}^{l-1} \mathbb{B}^B y_B.
\end{equation}
This can be independently derived from the observation that we can define the quantity
\begin{equation}
\mu_{tot} := \dfrac{\partial \mathcal{U}}{\partial n} \bigg|_{s,x_A} = \dfrac{\partial \mathcal{U}}{\partial n} \bigg|_{s,y_B} ,
\end{equation}
which can be considered a gauge-invariant (for chemical gauges of the kind \eqref{gaugeperfetto}) chemical potential of the fluid.

\section{Adiabatic expansion of ideal gases}\label{sec:adexp}

In this appendix we derive the formula for the transformation of the particle distribution function $f$ under an adiabatic expansion. Our derivation consists of two steps. 

Consider a single particle which is bouncing inside a cubic box of volume $V=L^3$. Suppose that the box is expanding infinitely slowly, then we have that
\begin{equation}\label{ADD}
\dfrac{dp_j}{p_j} = - \dfrac{dL}{L} ,
\end{equation}
where $p_j$ is the absolute value of the component $j$ of the momentum of the particle (the component of the momentum changes sign in any collision with the walls, therefore we consider only the absolute value). We give two proofs of this formula.\\

\textit{Proof from special relativity}

Let us study what happens when the particle collides with the moving wall which is normal to the direction $1$. Before the collision the four-momentum of the particle is $(\epsilon ,p^1,p^2,p^3)$. It is convenient to boost to the wall's frame, which is moving with infinitesimal velocity $w$. In this frame the four-momentum of the particle is
\begin{equation}
(\gamma \epsilon - \gamma w p^1, \gamma p^1 - \gamma w \epsilon, p^2,p^3 ),
\end{equation} 
with $\gamma=(1-w^2)^{-1/2}$. The walls are perfectly reflecting, therefore in this frame after the collision we have
\begin{equation}
(\gamma \epsilon - \gamma w p^1, -\gamma p^1 + \gamma w \epsilon, p^2,p^3 ).
\end{equation}
Transforming back to the original frame we find that the momentum after the collision is
\begin{equation}
p_1^{ac} = \gamma^2 (-p_1 + 2 w \epsilon  -w^2 p_1 )  \quad \quad p_2^{ac}=p_2  \quad \quad  p_3^{ac}=p_3.
\end{equation}
Considering that $w$ is infinitesimal we can neglect the order $w^2$, obtaining that 
\begin{equation}
p_1^{ac} = -(p_1 - 2w \epsilon),
\end{equation}
therefore the variation of the modulus of the momentum induced by a collision is
\begin{equation}
(dp)_{1coll} = -2w \epsilon.
\end{equation}
Assuming that only one of the two opposite walls is moving, then in a time $dt$ (sufficiently long to have many collisions with the wall, but sufficiently short to produce a small displacement of the wall) we have that 
\begin{equation}
dL = w dt.
\end{equation}
The number of collisions of the particle with the wall is
\begin{equation}
N_{coll} = \dfrac{v^1 dt}{2L} = \dfrac{v^1 dL}{2 w L},
\end{equation}
where $v^1$ is the absolute value of the the component $1$ of the velocity of the particle. The variation of the modulus of the momentum during $dt$ is
\begin{equation}
dp_1 = N_{coll} (dp)_{1coll} = -  \dfrac{\epsilon v^1 dL}{ L}.
\end{equation}
In the context of special relativity it must be true that $p_1 = \epsilon v^1$ both in the massive and massless case, so we find equation \eqref{ADD}.\\

\textit{Proof from quantum mechanics}

The Hamilton operator of the particle in the box has the form
\begin{equation}
\hat{H} = \epsilon(\hat{\vect{p}})+ V_{w}(\hat{\vect{x}}).
\end{equation}
$\epsilon$ is the kinetic energy and depends only on the modulus of the momentum, $V_w$ is the potential energy of the walls, it is zero inside the box and infinite outside. It is known that the eigenstates of such a Hamiltonian are the normal modes 
\begin{equation}
\psi_{\vect{a}} (\vect{x}) =   \prod_{j=1}^3 \sqrt{\dfrac{2}{L}} \sin \bigg( \dfrac{\pi a_j}{L} x^j \bigg)
\end{equation}
where $a_j \in \mathbb{N}$. The walls are located in $x^j =0$ and $x^j =L$. The quantities
\begin{equation}
p_j = \dfrac{\hbar \pi a_j}{L} ,
\end{equation}
can be interpreted as the average moduli of the corresponding components of the momenta and it is true that
\begin{equation}
\hat{H} \psi_{\vect{a}} = \epsilon(\vect{p}) \psi_{\vect{a}} . 
\end{equation}
Now, according to the adiabatic theorem, if the walls move infinitely slowly, then a particle which at the beginning of the expansion is occupying the eigenstate $\psi_{\vect{a}}$ at the end of the evolution will occupy the eigenstate $\psi_{\vect{a}}'$ associated to the final Hamiltonian. This means that if $L \rightarrow L'$, then
\begin{equation}
p_j = \dfrac{\hbar \pi a_j}{L}  \longrightarrow p_j' = \dfrac{\hbar \pi a_j}{L'},
\end{equation}
or, alternatively,
\begin{equation}
d (p_j L) =0,
\end{equation}
which is equivalent to \eqref{ADD}. 
\\

Now that we have proven how the momentum of a single particle in a box changes under an adiabatic expansion we can study the behaviour of $f$. Since we are supposing that during the process the two-body collisions do not have time to occur, the particles do not influence each other, therefore they evolve as they were alone in the box. This means that if in the expansion $L \longrightarrow \lambda L$, then, integrating \eqref{ADD}, the momentum of each particle transforms as follows:
\begin{equation}
\vect{p} \longrightarrow \Phi_\lambda(\vect{p}) = \dfrac{\vect{p}}{\lambda} .
\end{equation}
The last step consists of finding the transformation law of $f$. Let us consider an arbitrary volume $\Omega$ in the momentum space. Before the expansion there are in the box 
\begin{equation}
N(\Omega) =L^3 \int_\Omega f(\vect{p}) \, d_3 p
\end{equation}
particles in this volume. After the transformation there is the same amount of particles in the volume $\Phi_\lambda(\Omega)$, therefore
\begin{equation}
L^3 \int_\Omega f(\vect{p}) \, d_3 p = \lambda^3 L^3 \int_{\Phi_\lambda(\Omega)} f_\lambda (\vect{q}) \, d_3 q ,
\end{equation}
where $f_\lambda$ is the distribution after the expansion and we have used the fact that the final volume is $\lambda^3 L^3$. Now, making change of variables and using the fact that the Jacobian determinant of $\Phi_\lambda$ is
\begin{equation}
J_{\Phi_\lambda} = \dfrac{1}{\lambda^3} ,
\end{equation}
we find
\begin{equation}
 \int_\Omega f(\vect{p}) \, d_3 p =   \int_{\Omega} f_\lambda \bigg( \dfrac{\vect{p}}{\lambda} \bigg) d_3 p .
\end{equation}
Since this equality holds for any $\Omega$ we finally obtain our formula for the transformation of $f$ under adiabatic expansion:
\begin{equation}\label{transformation}
f_\lambda (\vect{p}) = f(\lambda \vect{p}).
\end{equation}
We conclude the appendix making a couple of fundamental coherence tests. We have shown that \eqref{comparison} always holds, therefore we need to check if it is satisfied by our formula for the adiabatic transformation. In the following we use a subscript $\lambda$ to denote the quantities obtained from the distribution $f_\lambda$, and we do not put any subscript for the quantities computed with $f$.

With a change of coordinate $\vect{q} = \lambda \vect{p}$ you can easily see from \eqref{densità} that
\begin{equation}\label{xcv}
n_\lambda = \dfrac{n}{\lambda^3} ,
\end{equation} 
from \eqref{energgia} that
\begin{equation}\label{xcvb}
\mathcal{U}_\lambda = \dfrac{1}{\lambda^3} \int \epsilon \bigg( \dfrac{\vect{q}}{\lambda} \bigg) f(\vect{q}) d_3 q
\end{equation}
and from \eqref{pressszione} that
\begin{equation}
\Psi_\lambda = \dfrac{1}{3 \lambda^4} \int q_j v^j \bigg( \dfrac{\vect{q}}{\lambda} \bigg) f(\vect{q}) d_3 q 
\end{equation}
The ratio between \eqref{xcvb} and \eqref{xcv} gives us
\begin{equation}
\tilde{\mathcal{U}}_\lambda = v \int \epsilon \bigg( \dfrac{\vect{q}}{\lambda} \bigg) f(\vect{q}) d_3 q .
\end{equation}
Taking the derivative with respect to $\lambda$ and remembering \eqref{VvvV} we find
\begin{equation}
\dfrac{d \tilde{\mathcal{U}}_\lambda}{d \lambda} = -3v \lambda^2 \Psi_\lambda,
\end{equation}
which, considering that $v_\lambda = v \lambda^3$, is equivalent to \eqref{comparison}.

The final coherence test consists of verifying that the entropy per particle is conserved in this transformation, proving that it is adiabatic. The first step of the proof consist of realising that the transformation \eqref{transformation} implies that, see \eqref{sigma},
\begin{equation}
\sigma_\lambda (\vect{p}) = \sigma (\lambda \vect{p}).
\end{equation}
Plugging this result in the formula for $s_\lambda$, see \eqref{SssS}, and changing variable in the usual way we get
\begin{equation}\label{qwertyuioopljhggf}
s_\lambda = \dfrac{s}{\lambda^3} .
\end{equation}
Taking the ratio with \eqref{xcv} we find that $x_s$ does not vary with $\lambda$, which is what we wanted to prove.

\bibliography{Biblio}

\label{lastpage}
\end{document}